\documentclass[journal]{IEEEtran}
\PassOptionsToPackage{bookmarks=false}{hyperref}
  \usepackage{times}
   
\usepackage{lineno,hyperref}

\usepackage{subfig}
\usepackage{graphicx}
\usepackage[export]{adjustbox}
\usepackage{textcomp}
\usepackage{multirow}
\usepackage{color}
\usepackage{cite}
\usepackage{amsmath}
\usepackage[table,xcdraw]{xcolor}
\usepackage{lipsum,multicol}

\usepackage[noend]{algpseudocode}
\usepackage{array}
\usepackage{enumerate}
\usepackage{afterpage}
\usepackage[linesnumbered,ruled,vlined,english,onelanguage]{algorithm2e}

\begin{document}
\begin{NoHyper}

\title{Smart Parking System Based on Bluetooth Low Energy Beacons with Particle Filtering}

\author{\IEEEauthorblockN{Andrew Mackey,~\IEEEmembership{Student Member,~IEEE}, Petros Spachos,~\IEEEmembership{Senior Member,~IEEE}, and Konstantinos~ N.~Plataniotis,~\IEEEmembership{Fellow,~IEEE}}

\thanks{This work was supported in part by the Natural Sciences and Engineering Research Council of Canada (NSERC). }
\thanks{ A. Mackey and P. Spachos are with the School of Engineering, University of Guelph, Guelph, ON, N1G 2W1, Canada. (e-mail:  \textit{mackeya}@uoguelph.ca, \textit{petros}@uoguelph.ca).}
\thanks{ K. N. Plataniotis is with the Department of Electrical and Computer Engineering, University of Toronto, Toronto, ON, M5S 3G4, Canada (e-mail: \textit{kostas}@ece.utoronto.ca).}

}

\maketitle

\begin{abstract}Urban centers and dense populations are expanding, hence, there is a growing demand for novel applications to aid in planning and optimization. In this work, a smart parking system that operates both indoor and outdoor is introduced. The system is based on Bluetooth Low Energy (BLE) beacons and uses particle filtering to improve its accuracy. Through simple BLE connectivity with smartphones, an intuitive parking system is designed and deployed. The proposed system pairs each spot with a unique BLE beacon, providing users with guidance to free parking spaces and a secure and automated payment scheme based on real-time usage of the parking space.  Three sets of experiments were conducted to examine different aspects of the system. A particle filter is implemented in order to increase the system performance and improve the credence of the results. Through extensive experimentation in both indoor and outdoor parking spaces, the system was able to correctly predict which spot the user has parked in, as well as estimate the distance of the user from the beacon. 
\end{abstract}


\IEEEpeerreviewmaketitle

\section{Introduction}
\IEEEPARstart{U}{rban} planning and smart cities are popular research fields, and the availability of wireless technology has been the major catalyst in its development. As populations grow in numbers and density, urban organization and optimization are more important than ever, especially in regards to congestion and space optimization. Specifically, vehicle parking is an aspect of urban development that could vastly benefit from the support of a smart and cost-efficient system. 

Current parking infrastructures do not provide people with a central  platform, nor do they provide real-time parking availability. Poor parking infrastructure creates a large list of problems for the city and its people. Drivers waste a significant amount of time searching for an available parking spot. This causes a lot of congestion on the road, poor utilization of available parking spaces, and can even contribute to added vehicle emissions. Furthermore, when a spot is found, it may not be in the most convenient location relative to the driver's desired locality. Another problem is that parking space efficiency may be low due to people not knowing of the availability. This is a result of people balking or passing by potential parking due to the impression that there is no availability. Finally, although some parking spots offer a check-in/out system to pay based on how long the spot was used, there is no universal parking system that offers the flexibility, security, and convenience of an automated, web-based payment scheme such as the one  proposed in this paper.  
 
Bluetooth Low Energy (BLE) beacons, usually referred to as beacons, can be exploited in order to implement a simple and intuitive parking system. Beacons are small transmitters developed for a plethora of Internet of Things (IoT) applications~\cite{beaconsforIOT}. They bring attention to their location by periodically transmitting data packets, containing a unique identification. The region or location can be determined by using Received Signal Strength Indicator (RSSI) techniques~\cite{sadowski}. Beacons are used rather than utilizing alternative technologies such as Wi-Fi for three key reasons; they are very cheap,  they are small in size, and they can be placed on top of any  existing infrastructure with ease. These make them a budget-conscious  solution that can be implemented easily in a smart city application such as smart parking.   

In this paper, a smart parking system based on BLE beacons is introduced.
The system can be accessed from the user's smartphone. All parking lots that have registered with the system would be able to provide real-time information about their available parking and pricing. Each spot in a given lot would have a uniquely identifiable beacon. Through a developed  beacon-based  Android application, users are able to locate the exact available parking spaces nearest to them that meet their requirements. Beacons  enable the guidance capabilities of the system, by providing their real-time availability. Once parked, the user is given a Uniform Resource Locator (URL) transmitted by the unique beacon of that spot. This is facilitated by Google's Eddystone beacon protocol~\cite{eddystonepacket}. The Eddystone protocol is used because it implements a specific packet type that stores an encoded URL. The closest beacon, that correlates to their parking spot is detected using the smartphone application. The user opens the URL displayed on their screen, delivered by the beacon and is subsequently brought to a secure server to register their vehicle information and pay for that spot. Once registered, a clock running at the server-side of the application, keeps track of the registration time and continuously calculates a rate until the user releases occupancy and vacates from the said spot. 
 
The major contributions of this work are listed below:

\begin{itemize}
\item We propose the use of particle filtering for parking spot estimation. The filter is calibrated, after a number of experiments, to meet the needs of the introduced system.

\item We design an Android application for the proposed system. The application collects the signals only from the beacons that are registered to the system. The collected values are forwarded to a server for further processing. 

\item We design a server application to provide useful information to the user regarding the parking availability in different areas and provide payment instruction to the user. The application also keeps useful analytics regarding the parking duration and can also charge the user based on the usage of the parking spot.  The user is guided to the payment website, automatically without the need to type in any information regarding the parking spot.
 
\item Through extensive experimentation in both indoor and outdoor environments, we demonstrate the efficiency of the proposed system and
the advantages of the usage of particle filtering.

\end{itemize}

The system reduces the time spent finding a spot,  simplifies payment infrastructure, as well as provides people with a flexible payment service that only charges based on the exact time used. Overall, such a system could greatly optimize the parking use in urban areas,  as well as provide people with a centralized and simple parking service. The novelty of this particular system is its automated payment scheme and parking availability estimation capabilities based on BLE beacon devices, facilitated by Google's Eddystone beacon protocol.

The rest of this paper is organized as follows; Section~\ref{related} details the related work regarding BLE beacons, localization, filtering techniques, and smart parking models. This is followed by the proposed architecture in Section~\ref{arch} and the parking availability estimation in Section~\ref{parking_section}. The experimental results  are presented in Section~\ref{experiment}, followed by the conclusions in Section~\ref{conclusion}. 
 
\section{Related Work}\label{related}

BLE beacons have attracted much attention as they are considered a promising direction for improving indoor localization~\cite{he, spachos1}.
Research in BLE localization attempts to make accuracy improvements on RSSI-based techniques through the use of filtering algorithms~\cite{mackey1}.  In~\cite{iBeaconProximity},  a server-based Kalman filter implementation is used to improve BLE proximity estimation. In~\cite{particleFilt}, the RSSI accuracy of a beacon-based micro-location system is enhanced through the use of a particle filter. Using Gimbal 21 series beacons, the optimal particle and beacon number selection are determined for the environment. This proved to be beneficial in improving the position accuracy of the presented localization system. In~\cite{GuassianKalman}, three filtering techniques: Kalman filter, Gaussian filter, and a hybrid of the two, applied to an indoor beacon-based positioning system in which three environments were tested. The hybrid Kalman-Gaussian Linear filter was determined to be the most effective at improving indoor location accuracy. In~\cite{globalsip},  the implementation of a fully mobile software-defined Kalman filter for BLE proximity estimation is presented. The paper concludes that the addition of a Kalman filter greatly improves the achievable accuracy for BLE-based proximity estimation.
  
In any system with wireless nodes, energy consumption is a critical characteristic that determines the lifespan of the system, as well as its scalability. BLE beacons were designed to use low power hence, it is important to understand the power characteristics of such devices on the market before designing a smart parking system. In~\cite{elsevier},  the power consumption between a solar-powered beacon and a simple button cell powered beacon device is compared. It is determined that the solar-powered beacon is more power efficient than its battery-powered counterpart, but it should be noted that this comes at the cost of transmission interval limitations. In~\cite{EnergyCompare}, it proves that BLE requires less energy than Wi-Fi. Finally, the work introduced in~\cite{BLE_EnrgyEstimation} explores the power consumption of BLE devices. The experiment is conducted using a star topology, where a smartphone is used as the receiving device. The results of this experiment indicate the level of energy consumption BLE-based systems require.

The idea of a smart parking system has been examined in recent years. An application growing in popularity is the PayByPhone mobile application~\cite{paybyphone}. The application works by prompting the user to enter the unique parking location code advertised by the specific parking space. The user then enters the amount of time they wish to park there and is free to add/extend their time on their phone from anywhere, provided they have a network connection. This application is now available in select countries and cities. Another smart parking system is proposed in~\cite{cloud_SPS}. The authors propose an RFID-based parking system where all users carry an RFID tag that is detected by a central node at each lot to be able to determine the number of vehicles in the lot. A cloud-based network allows users to reserve parking spaces and calculates a least cost metric as a function of distance and free spaces for the user to find optimal parking. In another study~\cite{Geng}, a smart parking system based on resource allocation and reservations is proposed. Utilizing smartphones for smart parking systems is also explored in~\cite{Bonde}, where Android smartphones are programmed to facilitate an automated parking system that requires little human interaction. In~\cite{Suryady} a cloud-based smart parking system rooted in IoT is proposed. The system is comprised of three layers, a Wireless Sensor Network  layer that utilizes ZigBee technology, an IoT middleware layer, and a front-end user interface layer. The ZigBee wireless nodes are placed so that there is a node under every vehicle in the lot, using ferromagnetic sensors to detect their presence.

All of these smart parking systems present their own advantages. However, for some of them, the user needs to register the correct parking spot~\cite{paybyphone}, have an online profile and interact with the application~\cite{Geng, Bonde} or carry special equipment~\cite{cloud_SPS} which is already registered in the system.  None have the capability of being cheap, easy to use and easily integrated with any infrastructure, all while simultaneously providing basic proximity and location services. This paper presents a smart parking solution based on BLE beacon which implements Google's Eddystone protocol to provide the user with payment information automatically, after parking the vehicle. The user needs to have a smartphone device with BLE capabilities and the developed Android application to access the system.

\section{Proposed Architecture}\label{arch}

In this section, a brief overview  introduces the system, followed by a detailed description of each technology, along with its functionality and performance enhancements within the system environment.

\begin{figure}[t!]
  \centering
  \includegraphics[width=0.9\columnwidth]{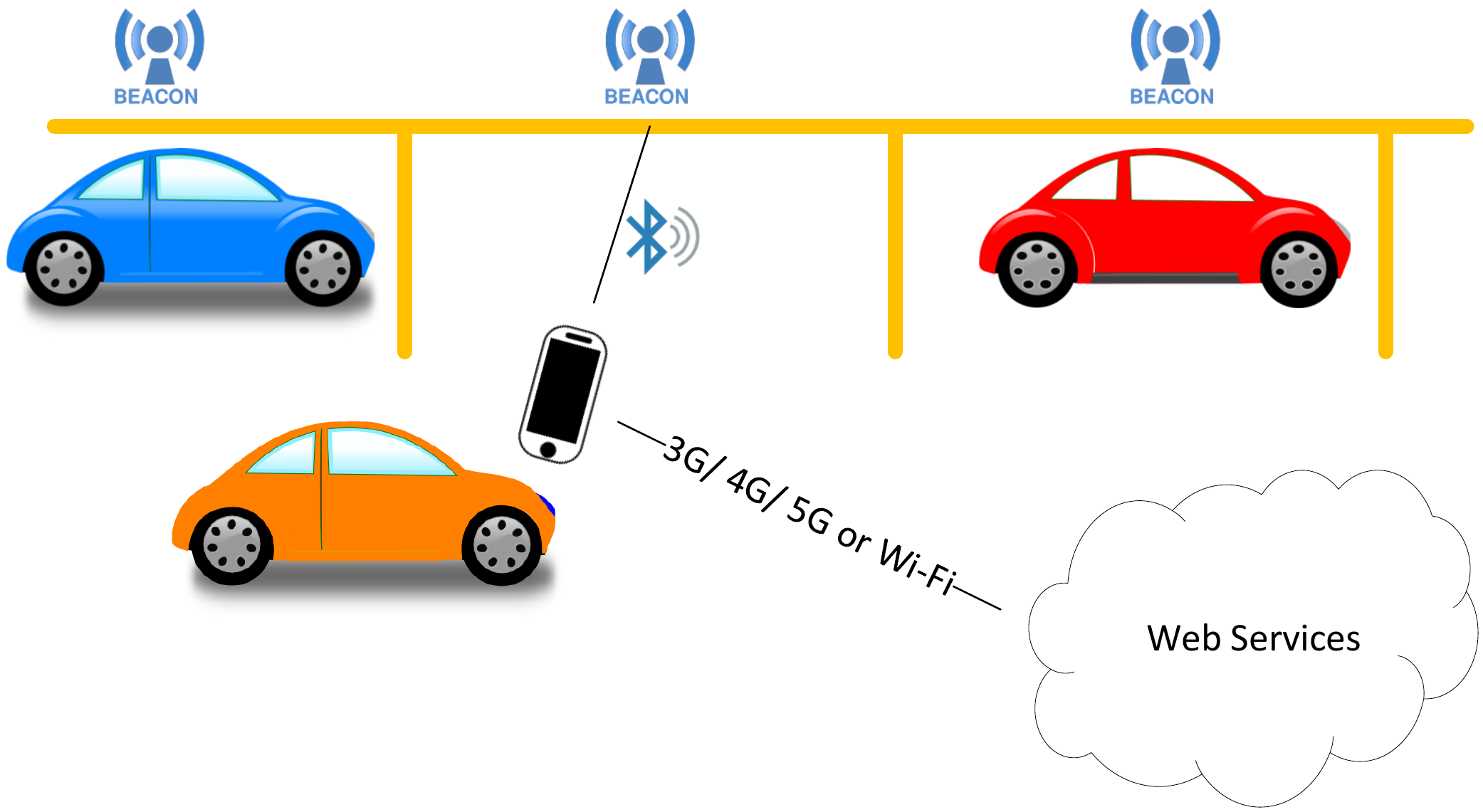}
  \caption{Smart parking system overview.}
  \label{overview}
\end{figure}

\subsection{System overview}
The proposed smart parking system relies on three main components that use different wireless technologies. A system overview is shown in Fig.~\ref{overview}.

\subsubsection{Cloud server}
The backbone of the system is the remote server and the cloud infrastructure. A remote server hosts all of the parking information, providing the application with parking spot location, availability status, associated costs, payment services, and management services. The web services provided by the remote server keep track of all parking information, its users, provide real-time status updates, handle payment, and offer to the registered parking providers with management services for its parking authority. All the communication between the web services and the user is over the smartphone application at the user's smartphone.

\subsubsection{Smartphone} The second critical component of the system is the smartphone. There are three features that the smartphone should provide:
 
\begin{enumerate}[i.]
\item Network connectivity
\item User profile creation
\item BLE connectivity
\end{enumerate}

Network connectivity is necessary for the user to go to the parking server that hosts the payment and parking lot information. This includes the ability to locate nearby parking spots and receive real-time availability status. 

For the user profile, the application supports the ability to save vehicle and payment information unique to the user. At the same time, the application keeps track of useful statistics for the user, such as preferable parking spots, and provides a recommendation, such as available parking spots close to events the user plan to attend.

The third necessary component is BLE connectivity. The majority of smartphones in today's market have the required hardware and capabilities to listen to and interact with BLE devices, specifically BLE beacons. However, in case BLE is not available, the users are able to enter the unique spot ID of each beacon, which is a letter to identify the lot, followed by a unique number to identify the spot.

\subsubsection{BLE beacons} The third critical component of the proposed smart parking system is BLE beacon devices. BLE beacons are small, low cost, and low power transmitting devices that implement Bluetooth Special Interest Group's Bluetooth Low Energy technology and was specifically designed for use in the IoT framework~\cite{beaconsforIOT, spachos1}. A beacon is placed at the side of each parking spot location, in a way that avoids physical tampering. This could be a location near the ceiling in a parking garage or locked behind a protective covering/shield on the parking curb in other lot configurations. Each BLE beacon can be configured to have a unique identifier, which must be configured during deployment and provided to the remote server/database. Using proximity estimation, the smartphone of any given user is able to detect the beacon that is closest to it, giving users the ability to register their vehicle with the corresponding parking spot. 

The flow chart of the system can be seen in Fig.~\ref{userFlow}. It depicts the sequence of events for a standard user. The user finds a parking region, finds a spot and parks, scans for the closest beacon corresponding to their spot, obtains the URL from the beacon and registers for the spot via a secure web server. The necessary information is stored/tracked for parking management. When the user returns to their vehicle, all they have to do is unregister from the spot and payment is automatically processed. To avoid overpayment, users can either pre-pay for the parking spot or set maximum parking time limits within the application/registration process. 

\begin{figure}[t!]
  \centering
  \includegraphics[width=0.9\columnwidth]{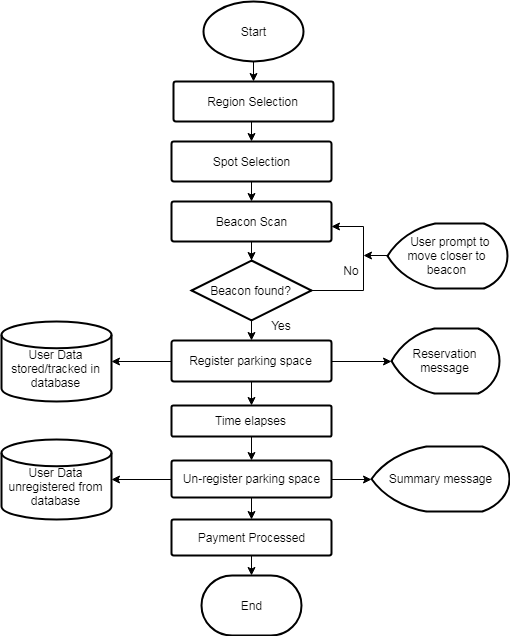}
  \caption{Flowchart of the introduced system.}
  \label{userFlow}
\end{figure}

\subsection{System implementation}\label{hardware}

\subsubsection{Server requirements and development}
 The server manages all data relating to the parking lot with an emphasis on security. This dedicated server must not be restricted by a Wi-Fi network, meaning that the end-user should have access when using mobile data. Although one could argue a company parking lot should restrict the application to one network, this may not always be the most convenient for the user.

At the same time, the server must be able to handle all errors gracefully. This includes any time a user tries to take a spot already in use as well as somebody forgetting to sign out of the spot. In terms of transaction handling, the server will collect the user's info and ensure that the credit card information is valid. If a transaction cannot be made, the server  notifies the user as soon as possible and the spot will be considered illegally parked until the user is able to pay the fee or the spot is vacated. The server is not responsible for handling illegally parked vehicles, it will only be able to alert the system administrator.

\subsubsection{Mobile application}
 For the introduced smart parking system, the application should be able to support cloud connectivity, user profiling, payment services, and a simple user interface. To collect the raw data from the BLE beacons, \texttt{AltBeacon} library~\cite{altbeacon}  was used. The library was used only for the data collection, while for distance estimation, we used our formulas and libraries. In the introduced system, the mobile application is developed in Android OS. Specifically, the application is developed for a Google Nexus 5 running Android 6.0.1. For an iOS application, there is no need for the \texttt{Altbeacon} library, since an iBeacon library already exists in the system. The application is comprised of five main activities:

\begin{enumerate}[i.] 
\item Locating nearby parking
\item Profiling user and vehicle information
\item Setting up the payment information
\item Parking registration capabilities
\item Find my car capabilities
\end{enumerate}

Three of the activities are shown in Fig.~\ref{mobileapp}. A simple menu bar would allow the user to switch between activities, as seen in Fig.~\ref{menuScreen}. In Fig.~\ref{userInfo}, it is shown what the  user/vehicle information screen looks like. The user needs to have the application running in the background while looking for parking or start  the application after parking the car. After parking the car, the user scans for the beacon, and its associated spot, by hitting the radar floating action button, shown in Fig.~\ref{home}. The smartphone detects the closest beacon and registers the user to the associated parking space. The user does not need to take any further actions about the spot.

The payment information is another activity, where the user can set up a payment method. The last activity is the  ``Find my car",  in which the application shows the user the path between the smartphone location and the last registered parking spot.

 When discovering and registering for a specific spot, a simple graphic is displayed to the user, as shown in Fig.~\ref{lot}, that depicts the parking lot in Fig.~\ref{realpark}, and where the X represents unavailable spots, while the checkmark represents the spot registered to the user's car, while the rest of the spots are available.

In addition to the application development and services, the smartphone should  be able to interact with  beacons and access this information within the smart parking application. Hence, the mobile application turns on the Bluetooth at the smartphone to listen to nearby beacons. The application displays and utilizes only the nearest beacons corresponding to the desired spot, which are registered within the smart parking system and not any nearby BLE device. 

\begin{figure}[t!]
\centering
\captionsetup[subfloat]{farskip=0pt}%
\subfloat[Menu screen.]{\includegraphics[width=0.31\linewidth]{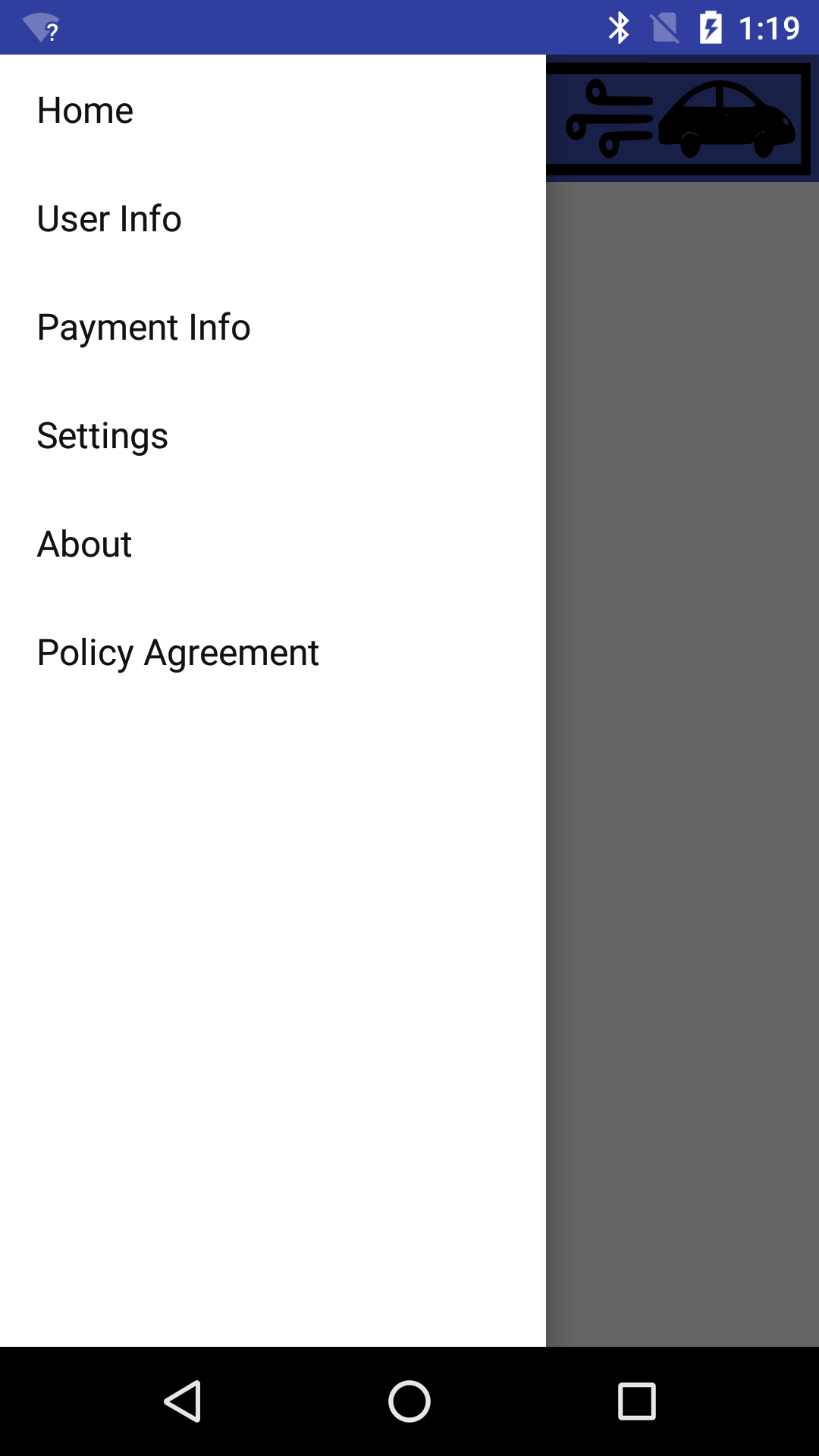}
\label{menuScreen}}\hfill
\subfloat[User info screen.]{\includegraphics[width=0.31\linewidth]{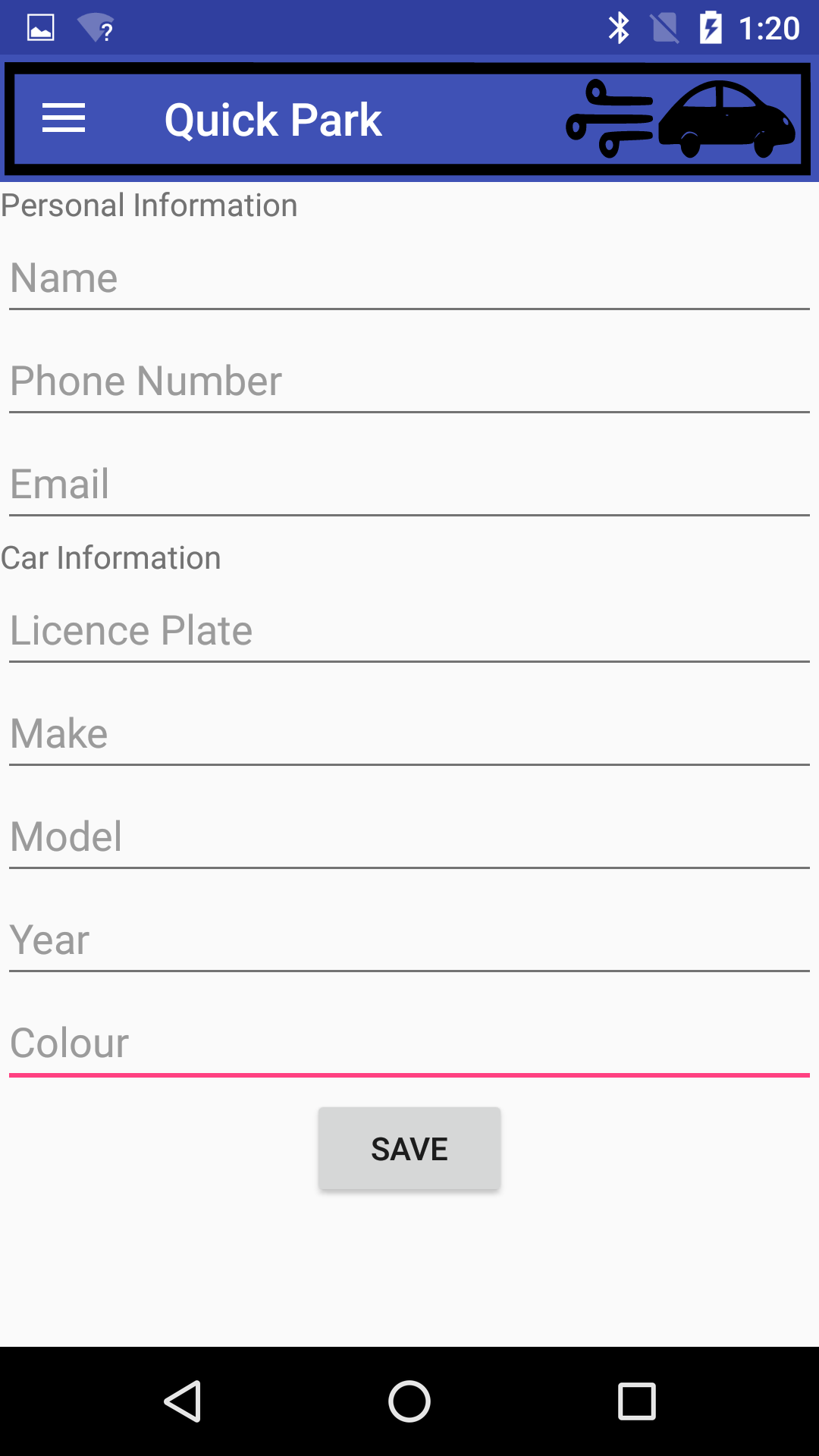}
\label{userInfo}}
\label{fig:hw1}
\subfloat[Parking spot scan screen.]{\includegraphics[width=0.31\linewidth]{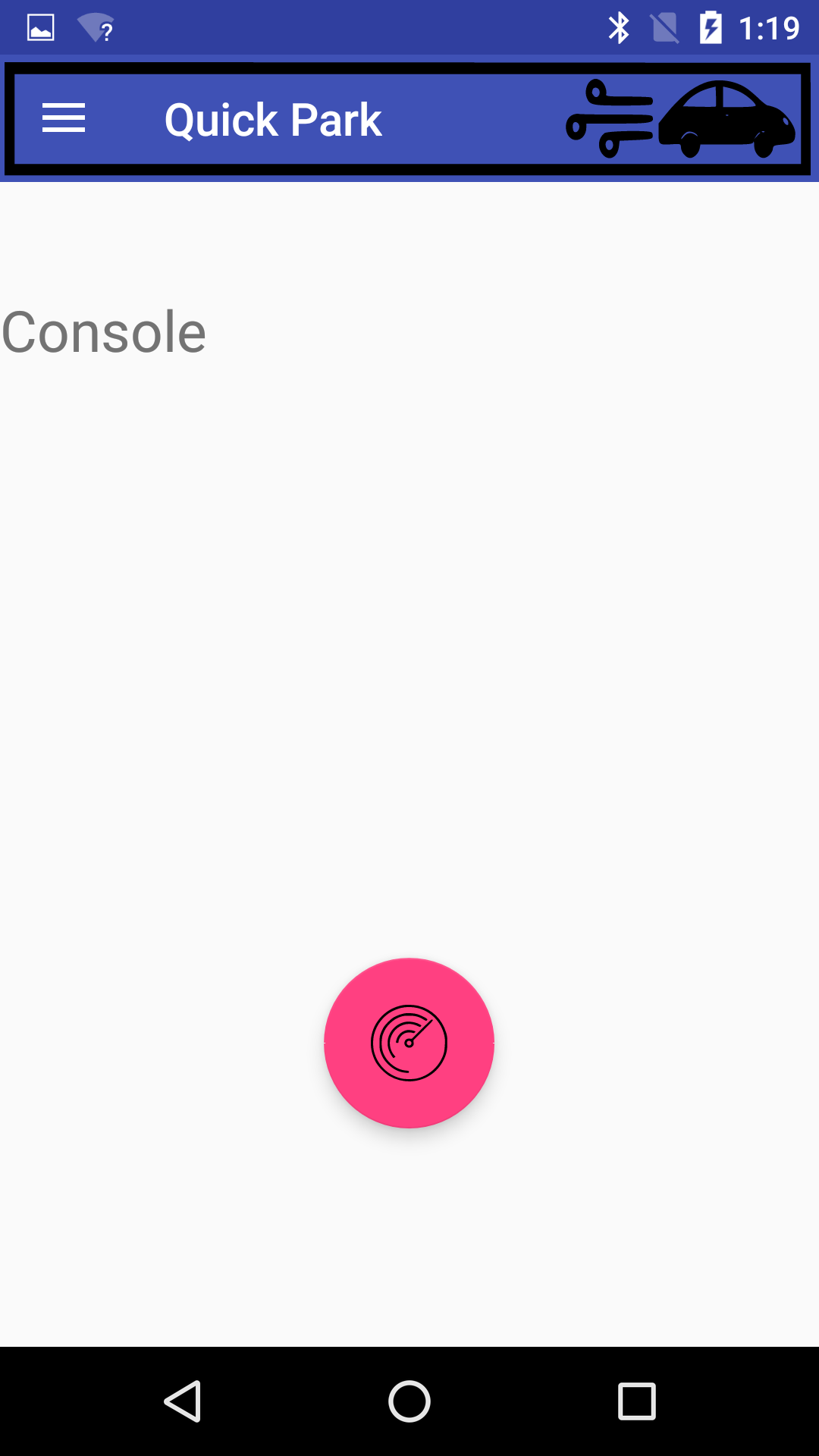}
\label{home}}
\label{fig:hw1}\hfill

\caption{Developed mobile application for the system.}
\label{mobileapp}
\end{figure}

\begin{figure}[t!]
\centering
\captionsetup[subfloat]{farskip=0pt}%
\subfloat[Availability screen.]{\includegraphics[width=0.39\linewidth]{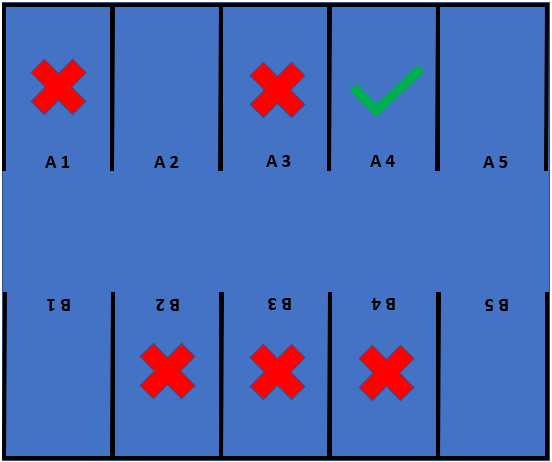}
\label{lot}}
\subfloat[Parking lot.]{\includegraphics[width=0.6\linewidth]{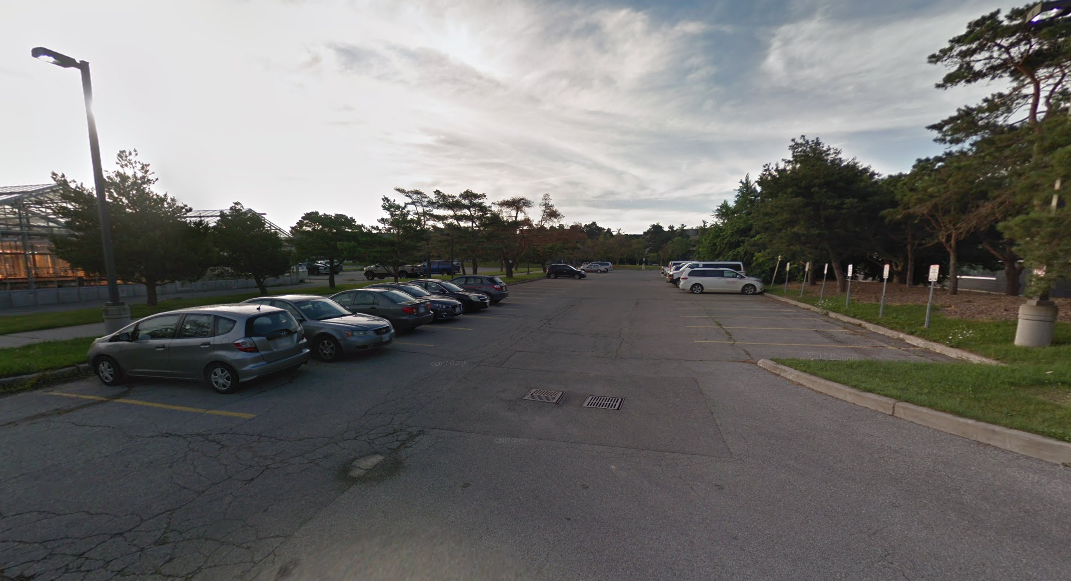}
\label{realpark}}
\caption{Parking lot for experiments with the introduced system.}
\label{parkingfig}
\end{figure}

\begin{figure}[t!]
\centering
\captionsetup[subfloat]{farskip=0pt}%
\subfloat[Gimbal Series 21.]{\includegraphics[width=0.29\linewidth]{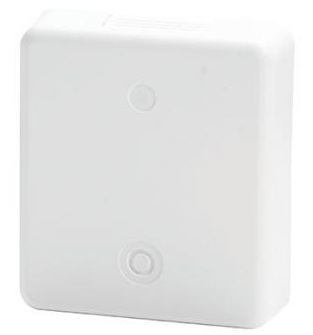}
\label{gimbal}}\hfill
\subfloat[BLE beacon at the parking spot.]{\includegraphics[width=0.6\linewidth]{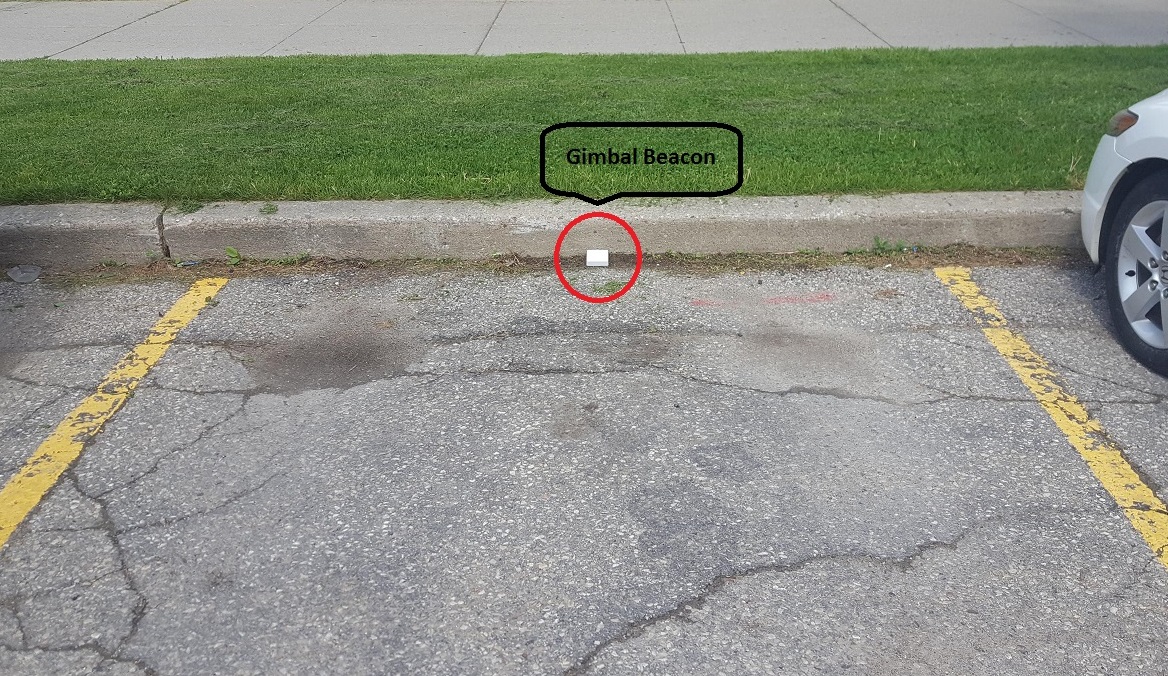}
\label{gimbalpark}}
\caption{BLE beacons used at the smart parking system.}
\label{parkinglot}
\end{figure}

\subsubsection{BLE beacons}\label{beacons}

Gimbal Series~21 beacons, shown in Fig.~\ref{gimbal}, are used and deployed at the parking lot, as shown in Fig.~\ref{gimbalpark}. Series~21  operates on 4~AA batteries which make the battery life superior to common button cell powered beacons. It has dimensions of 86 mm$\times$77 mm$\times$25 mm. It is configured to implement Google's Eddystone protocol~\cite{eddystonepacket}. 
	
Series 21 was specifically chosen due to its wide availability, extended lifespan over other beacon models, and simple configuration/management software. Beacon devices are not able to receive any information from the user smartphone, it only periodically transmits its unique ID. This drastically reduces the privacy concerns and likely leading to wide user acceptance. 

\subsection{Wireless communication}\label{connectivity}
The different components communicate wirelessly, between the smartphone and the BLE beacons as well as the smartphone and the server. 

\subsubsection{BLE communication}\label{BLE_COMMS} 
BLE beacons are capable of implementing their own transmission protocol. In this work, Google's Eddystone beacon protocol was used~\cite{eddystonepacket}.  Eddystone defines 3 frame types: UID, URL, TLM. 
\begin{itemize}
\item UID packet transmits the unique identifiers. It can be used to identify a unique region and parking spot. In reference to Fig.~\ref{lot}, the UID contains data that identifies the specific lot, section (A or B), and unique spaces (1 through 5). 

\item URL packet broadcasts a compressed URL that can be used by the receiving client. 

\item TLM packet allows for the transfer of any telemetry/sensor data. 
 \end{itemize}
 
There is also the EID frame which is a time-varying packet structure that provides added security~\cite{eddystonepacket}. Due to Eddystone's pre-defined URL packet configuration, it is an excellent choice for implementing the introduced parking application. 	

BLE beacons also offer two other configurable parameters, the transmission interval and the transmission power. The transmission interval defines how often the beacon transmits its identifiers. An increase in transmission interval frequency will drastically affect the battery life of the beacon. The transmission power increases the possible transmission range, i.e. it changes how far the beacon can be seen from~\cite{elsevier}. This also reduces the battery life of the beacon if it increased. To reduce interference among beacons and increase the lifespan of the proposed smart parking system, the transmission power and interval should be set relatively low and after experimentation at the deployment area. 

\subsubsection{Cloud communication}
A remote server is used to simulate and manage many aspects of the system.  This cloud-based implementation is communicated with the client Android application using a simple Transmission Control Protocol (TCP) connection over either LTE or Wi-Fi. When a user has found a spot and corresponding Bluetooth beacon, they may check the status of the spot by passing the beacon information back to the server to confirm the availability and view the cost. Next, they may choose to register to that spot by clicking on the URL packet that is given to them, which will link the user to the parking space and automatically pass their profile and payment information to a secure server. The parking time is tracked all the time and  the total cost is calculated via the server to ensure an accurate measurement. Once unregistered, the calculated cost is  charged to the user via their credit card. 

A screenshot of the server running is shown  in Fig. ~\ref{server}. When the server script starts, a list of available spots is displayed. An event in the system shows the type and an updated list of spots that separates those available and unavailable. The list of taken spots has all the user information for each spot, making it easy for any lot manager to keep track of users.

\begin{figure}[t]
\centering%
\includegraphics[width=0.9\columnwidth]{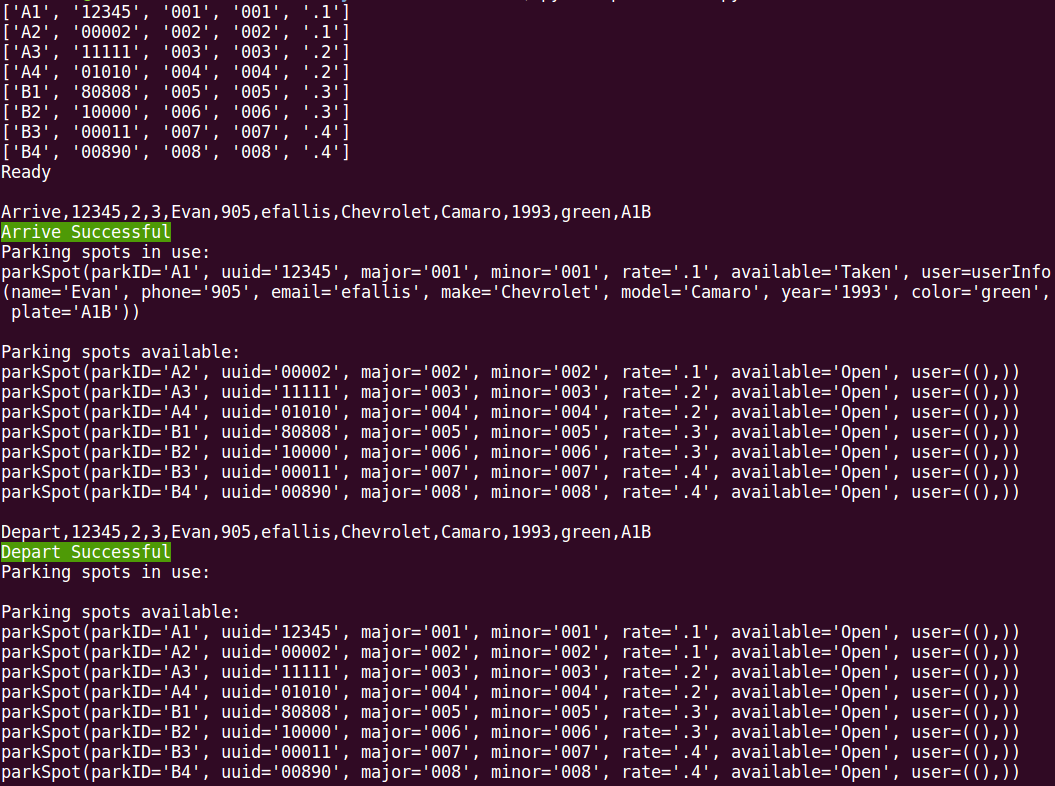}%
\caption{Server screen.}%
\label{server}%
\end{figure} 

\section{Parking Availability Estimation}\label{parking_section}
When a car approaches an available parking spot, the smartphone starts receiving signals from the beacons around the parking spot. To improve the accuracy of these signals, a number of filtering techniques are applied.

\subsection{Averaging}\label{avg}
In order to smooth the results and eliminate outliers in RSSI measurements, a simple averaging algorithm is applied to the raw data. Specifically, the averaging interval is done over the entirety of the data collection at a given distance. For instance, at a transmission interval of 1 second and a measurement period of 1 minute, it can be expected that 60 values are to be averaged using the following formula:
    
\begin{equation}\label{avgEqn}
RSSI_{avg} = \frac{\sum_1^n{RSSI_i}}{n}
\end{equation}
where $n$ is the number of the $RSSI$ values.

\subsection{Particle filtering}\label{filter}
The particle filter is a nonlinear, Bayesian filter often used in signal processing, especially when observations/ measurements are noisy~\cite{mackey1}. Since proximity estimation using beacons is a noisy non-linear Bayesian problem, the particle filter is a great suitor. It was specifically chosen over other filter implementations such as the Kalman or Gaussian filters. Kalman and Gaussian filters are simple to implement on smartphones and have been explored in-depth with regards to BLE beacon distance approximation~\cite{globalsip, GuassianKalman}.  However, they rely on the linear- Gaussian assumption, which is often violated in real experiments due to the nonlinear relationship between RSSI measurements and the position. Also, since the proposed system model makes use of a server for parking management services, the server can also be exploited in order to implement computationally expensive filters like the particle filter.

The particle filter estimates the current state by utilizing all measurement observations $k$ up to and including time $i$. The particle filter is designed and laid out similar to~\cite{particleFilt}. The measurement is represented by a state sequence $\{x_i, i \in N\}$ given by: 

\begin{equation}\label{stateSeq}
x_i = g_i(x_{i-1}|m_{i-1})
\end{equation}
where $g_i$ is a non-linear state function of the previous state $x_{i-1}$ and  $\{m_{i-1}, i \in N\}$ is an independent, identically distributed process, representing the noise sequence.

The particle filter estimates the current state $x_i$ recursively by utilizing all measurement observations $k_i$ up to and including time $i$. The computational load of each measurement is given by:

\begin{equation}
k_i = h_i(x_i|n_i)
\end{equation}
where $h_i$ is a non-linear function of the current state $y_i$ and the independent measurement noise sequence $n_i$ of the system.

Let $\{x_{0:i}^j, w_{i}^j\}_{j=1}^{N_s}$  denote a set of  random measurements to characterizes the posterior pdf $p(x_{0:i}|k_{1:i})$, where
$\{x_{0:i}^j, j=0,...,N_s\}$ is the support points set which has associated weights $\{w_i^j, j=1,...,N_s\}$, and $x_{0:i}=\{x_n, n= 0,...,i\}$ is the set of all states up to time $i$. The weights are all normalized at 1 and the posterior density at $i$ is approximately:
\begin{equation}
p(x_{0:i}|k_{1:i}) \approx \sum_{j=1}^{N_{s}}{w_i^j}\delta(x_{0:i} - x_{0:i}^j)
\end{equation} 

This is a discrete weighted approximation of the posterior probability. Although the samples are chosen according to their significancy on the global density, we can not concentrate the samples adaptively on the high probability regions in order to obtain variable resolution. Particle filter takes advantage of Monte Carlo numerical integration methods to provide a recursive implementation of the Bayes filter~\cite{yin, mohammadi, papi}. The weights are chosen using importance sampling, and the weighted approximation density is:
\begin{equation}
p(x) \approx  \sum_{i=1}^{N_{s}}{w^j}\delta(x - x^j)
\end{equation}
where
\begin{equation}
w(x_j) \propto \frac{\pi(x^j)}{q(x^j)}
\end{equation}
is the normalized weight of the $j$th particle.

The principle of the importance sampling is used to approximate the posterior distribution as~\cite{particleFilt}:

\begin{equation}
p(x_i|k_{1:i}) \approx \sum_{j=1}^{N_{s}}{w_i^j}\delta(x_i - x_i^j)
\end{equation}

where the weights are:
\begin{equation}
w_i^j \propto  w_{i-1}^j \frac{p(k_i|x_i^j)p(x_i^j|x_{i-1}^j)}{q(x_i^j|x_{i-1}^j), k_i}
\end{equation}

Hence, the SIS algorithm consists of recursive propagation of the weights, since each measurement is received sequentially.

\subsection{Particle filtering implementation}
\par The generic particle filter algorithm is executed in five steps:

\begin{enumerate}[1)]
\item Generate particles
\item Prediction
\item Update particle weights
\item Resample
\item Estimate
\end{enumerate}

The particle filter designed for this experiment is only concerned with one dimension, hence the proximity distance measurements. Thus, the particles generated incorporate only a single state, which is the distance in meters as a function of RSSI.

The parameters of the experiments are only concerned with measurements between 0 and 4 meters. Hence, the particles generated are random values between the intervals of 0 and 4. All particles are initialized to have equivalent weights (probabilities) $1/N$, where $N$ is the number of particles. The prediction step is supposed to alter the particle state by some form of control input. This would be equivalent to user movement towards and away from a particular BLE beacon. Since the experiments are only concerned with building a path loss model with static proximity measurements, the prediction step is left out of the algorithm. 

The next step is to update the weights corresponding to all of the particles. This is done by computing how close each particle is to the current measurement value. Particles which are closer to the measured value receive a larger weighting. The weight update is computed as a 1-dimensional Gaussian fit that accounts for measurement error. 

First, the absolute difference each particle is to the current measurement computed and squared:
\begin{equation}\label{updateWeights1}
x_j = |particle_j - z_i|^2
\end{equation}
where $x_j$ is the relative closeness the particle is to the measurement, with  $j$ ranging from 1 to $N$.

Then, a gain factor is computed based on a Gaussian distribution that updates the effective weight of that particle: 

\begin{equation}\label{updateWeights2}
k_j = e^{\frac{-0.5*x_j}{n_i^2}}
\end{equation}
\begin{equation}\label{updateWeights3}
weights_j = weights_j*k_j
\end{equation}
where $k_j$ is the calculated gain factor and it is a function of $x_j$ and the measurement noise $n_i$.

This will give larger weights to those values that are closer to the actual measurement. Finally, the weights are all normalized about 1. A measurement error of 1.2~m was found to be optimal for this step, as determined by iterative trials of the particle filter.

 After the weights are normalized, the probability density of the particle set is updated by the newly calculated weights.	
 
 \begin{algorithm}[t]
\KwResult{Updated particles and weights arrays}
 \eIf{$N_{eff} < N/2$}
 {
   Get cumulative sum array of particle weights\\
   Generate array (length of weights) of random numbers between 0 and 1\\
   Sort the cumulative sum by the random number array\\
   Find the indices of the weights that match\\
   Update the particles by replacing the unfit ones with new better fitting values\\
   Update the weights of the new particles to be once again equal\\
 }{Continue}
 \caption{Multinomial Resampling}
 \label{resampleAlgo}
\end{algorithm}

This is followed by resampling. Particle filter suffers from degeneracy phenomenon, since after a few iterations of the algorithm, all the particles are assigned negligible weights, except for one~\cite{pak,xie}. This effect can be reduced by the choice of  importance sampling function and resampling. During resampling, improbable or unfit particles are discarded in favor of new, more probable particles. The multinomial resampling method, described by Algorithm~\ref{resampleAlgo}, is used. Resampling is not conducted in every iteration of the particle filter loop. Only when the particle filter degenerates, i.e. the number of useful particles falls below a threshold, is the resampling step executed. In order to determine when to resample, the number of effective particles is calculated as the inverse sum of squares of all the weights, shown  below: 

\begin{equation}\label{neff}
N_{eff} = \frac{1}{\sum{weigths^2}}
\end{equation}

When this value falls below $N*\beta$, where  $\beta$ is a scaling factor between 0 and 1, then the resampling procedure is executed. 

Finally, the filter can estimate the new state by calculating a discrete weighted approximation, i.e. the weighted mean by:

\begin{equation}\label{w_mean}
w_{mean} = \frac{\sum{weights * particles}}{\sum{weights}}
\end{equation}

 It can also give a confidence to the state estimation, calculating the weighted standard deviation of all the particles:
 
\begin{equation}\label{w_std}
w_{std} = \sqrt{\frac{\sum_{i=1}^n{weights_i*(particles_i - \mu)^2}}{\frac{(N'-1)}{N'}*\sum_{i=1}^n{weights_i}}}
\end{equation}
where $\mu$ is the mean of the weights and $N'$ is the number of non-zero weights.

 A $\beta$ value of 0.5 was found to be optimal for this experiment, with the 1000 particles. This number was determined through iterative trials, shown in the following section.

\section{Experimental Procedure and Results}\label{experiment}

In this section, the experimental procedure is presented followed by the  description of each experiment and a discussion on the results.

\subsection{Experimental procedure}
In order to evaluate the performance of the introduced systems, three experiments were conducted:
\begin{itemize}
\item \textbf{Path loss model experiment}, to build the path loss model for each of the environments. The path loss experiment conducted before the other two experiments.
\item \textbf{Distance estimation experiment}, to calculate the accuracy and the precision of the BLE beacons in each environment.
\item \textbf{Proximity identification experiment}, to examine the performance of the BLE beacons in terms of proximity estimation when they are used in a smart parking scenario.
\end{itemize}

	All the experiments took place in two parking environments, one indoor and one outdoor. Each beacon was placed at a base reference point, which was the central base of the desired parking spot. The beacons are first configured to have a transmission power of 0 dBm and a transmission interval of 1 second for the purposes of the experiments. The 0 dBm transmission power is set in order to achieve high accuracy, while the transmission interval is decided after experimentation, in order to obtain enough data points over time to  perform the filtering.

\subsection{Path loss model}

In order to be able to implement a more accurate distance estimation and proximity identification system, using RSSI techniques with BLE beacons and smartphones, a path loss model was developed. Once the path loss model is obtained, it can be used to characterize efficiently the behavior of the RSSI at different distances in the specific environment. Using a path loss model that reflects the characteristics of the environment can improve the estimation accuracy of the system.

 To obtain the path loss model for each environment, one BLE beacon was placed in a fixed position and RSSI measurements were taken on a user device over a period of 5 minutes.  During the experiment, people and other objects could move freely between the beacon and the smartphone, creating a combination of Line Of Sight (LOS) and Non-Line-Of-Sight (NLOS) measurements. This approach was followed to build a realistic path loss model. The distance between the BLE beacon and the smartphone was  20~cm and it was increased after every interval by 20~cm, up to and including 4~m, for both environments. The average RSSI along with the standard deviation for each fixed position is shown in Fig.~\ref{rssi_all}.

   \begin{figure}[t!]
\centering
\captionsetup[subfloat]{farskip=4pt}%
\subfloat[Indoor environment.]{\includegraphics[width=0.9\linewidth]{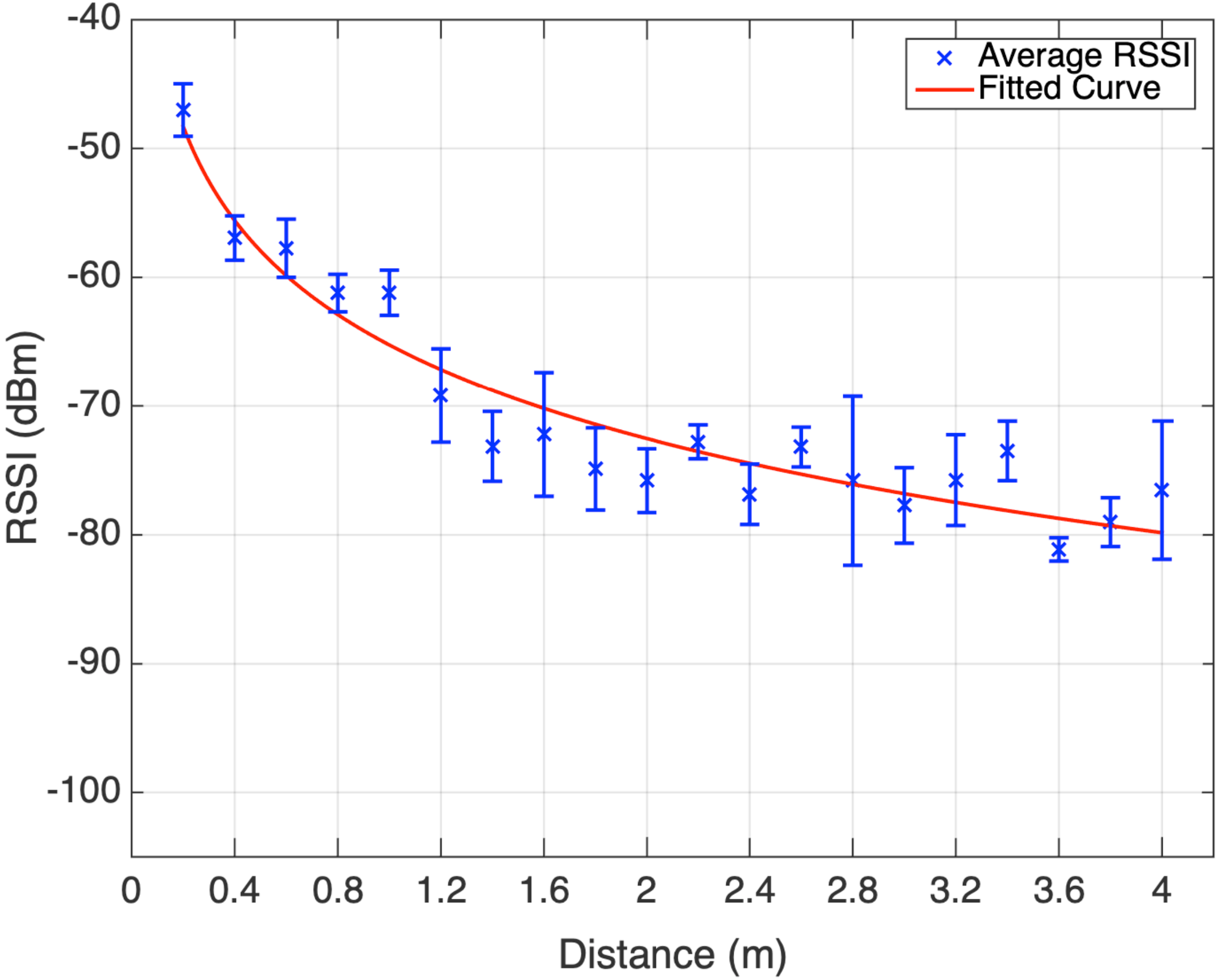}
\label{curveFit_indoor}}

\subfloat[Outdoor environment.]{\includegraphics[width=0.9\linewidth]{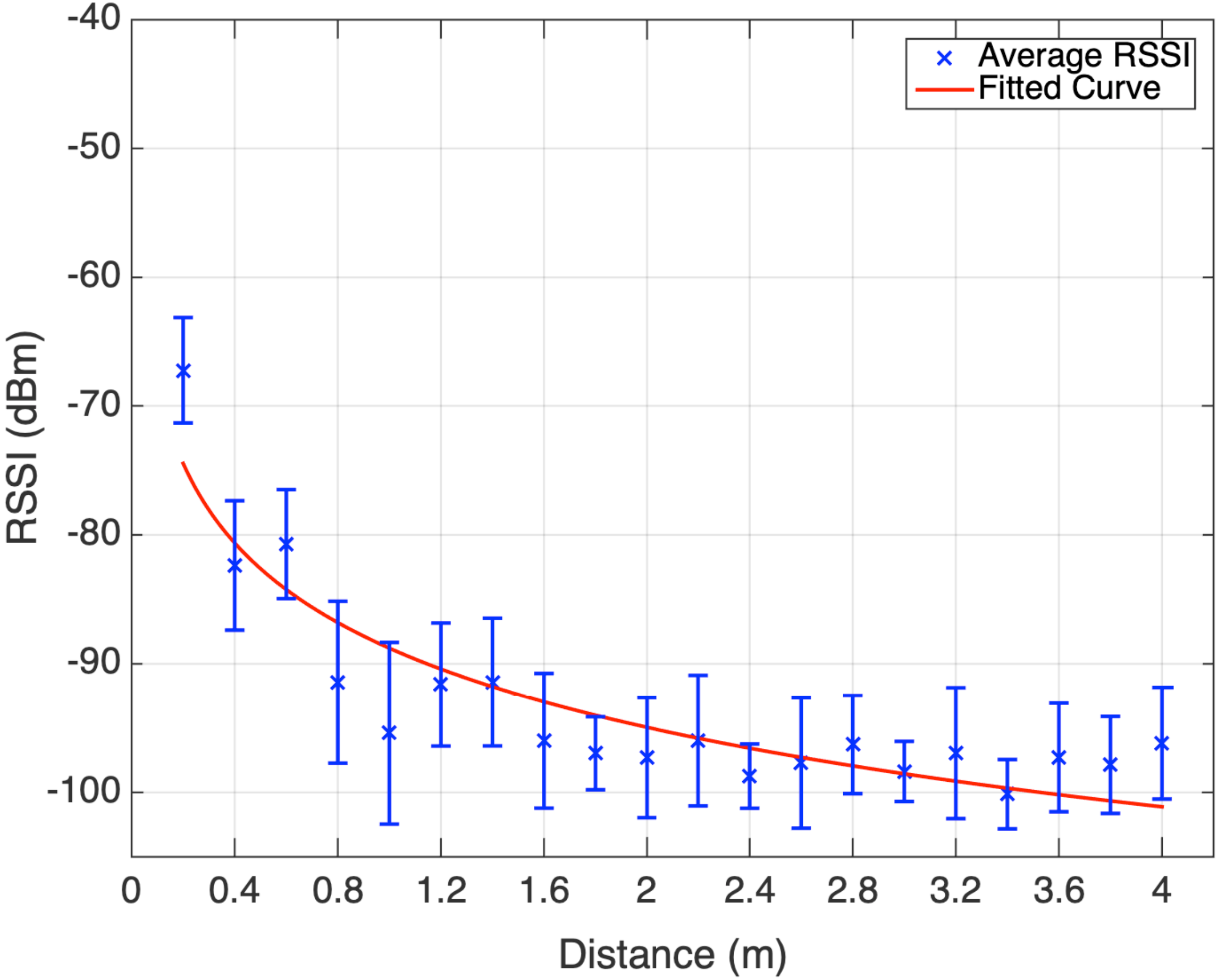}
\label{curveFit_outdoor}}
\caption{Best curve fitting for RSSI values at distances from 0.2 to 4 meters \protect\subref{curveFit_indoor} at the  indoor and \protect\subref{curveFit_outdoor} at the  outdoor parking.}
\label{rssi_all}
\end{figure}

The average RSSI value can be used to calculate the distance between the beacon and the device, through the log-normal shadowing model~\cite{ pathlossEstimationRSSI}, as follows:

\begin{equation}\label{pathlossEqn}
RSSI = -10\ n\ log_{10} (d/d_0)+C
\end{equation}
where $n$ represents the path loss exponent that varies depending on the conditions of the surrounding environment, $d$ represents the distance between the beacon and smartphone, $d_0$ is the reference distance which is 1~m in our case, and $C$ is the average RSSI value at $d_0$.

Next, knowing that the relationship between the distance and the RSSI is represented by Eq. (\ref{pathlossEqn}),  Matlab's curve fitting function was used, to estimate a curve for distance versus RSSI and to solve for our $n$ and $C$ parameters shown by the path loss function.

The curve fit line for each environment is shown in Fig.~\ref{rssi_all}.  As a result, the path loss factor $n$ is calculated to be $n=2.424$ with 95\% confidence bound between (2.058, 2.79) for the indoor environment, shown in Fig.~\ref{curveFit_indoor}, and $n=2.049$  with 95\% confidence bound between (1.581, 2.517) for the outdoor parking lot, shown in Fig.~\ref{curveFit_outdoor}. This is expected, as the free space path loss exponent is known to be $2$~\cite{freeSpace_PLE}. Similarly,  $C$ is calculated to be $C=-65.24$  with 95\% confidence bound between ($-66.74, -63.74$)  and $C=-88.78$ with 95\% confidence bound between ($-90.7, -86.87$), for the indoor and outdoor environments, respectively. 
In rearranging the path loss formula from Eq. (\ref{pathlossEqn}), we can estimate the distance using the following equation:
    
\begin{equation}\label{distanceEqn}
d = 10^{\frac{C-RSSI}{10\ n}}
\end{equation}

Hence, Eq. (\ref{pathlossEqn}), is rearranged into:

\begin{equation}\label{pathlossEqnin}
RSSI_{indoor} = -10\times 2.424 \times\ log_{10} (d) -65.24
\end{equation}
\begin{equation}\label{pathlossEqnin}
d= 10^{\frac{65.24+RSSI_{indoor}}{-24.24}}
\end{equation}

for the indoor environments, while:

\begin{equation}\label{pathlossEqnout}
RSSI_{outdoor} = -10\times 2.049 \times\ log_{10} (d) -88.78
\end{equation}
\begin{equation}\label{pathlossEqnout}
d= 10^{\frac{88.78+RSSI_{outdoor}}{-20.49}}
\end{equation}
for the outdoor environment.

The BLE beacon packet contains the suggested $C$ parameter, equals to $-62$, which is the factory-calibrated, read-only constant and indicates the expected RSSI at a distance of 1~m from the beacon. With respect to the outdoor environment, this value is far from the expected. This may be attributed to the more dynamic outdoor environment compared to indoors. In any case, it is clear that calibration and path loss experimentation is necessary before the system deployment.

\subsection{Distance estimation}
Once the path loss model was developed, the distance estimation experiment took place. In this experiment, the smartphone is placed at 8 locations from the beacon; 0.5~m, 1~m, 1.5~m, 2~m, 2.5~m, 3~m, 3.5~m, and 4~m. In each location, the smartphone collected data for approximately 2 minutes, while the experiment was repeated three times in each environment.  

The main limitation of the particle filter relates to its computational complexity. To find the optimal number of particles, we varied the number from 200 up to 2000, using an increment of 200 particles. We did not increase the number of particles more, due to smartphone energy and processing constraints. The error in meters, the MSE, and the standard deviation are shown in Table~\ref{table_par}.  The optimum performance for the first 2 meters is obtained with 1000 particles, while after the 2 meters the error is too high. For the outdoor environment, the results are similar, hence, 1000 particles were selected as the optimal number.

  \begin{table*}[t!]
\centering
\begin{tabular}{|c|c|c|c|c|c|c|c|c|c|c|c|c|}
\hline
\multirow{ 3}{*}{Particles} & \multicolumn{ 12}{c|}{Distance (m)} \\\cline{2-13} 
&\multicolumn{ 3}{c|}{0.5}  & \multicolumn{ 3}{c|}{1} & \multicolumn{ 3}{c|}{1.5} & \multicolumn{ 3}{c|}{2} \\ \cline{2-13}  
& Error & MSE & $\sigma$ & Error & MSE & $\sigma$& Error & MSE & $\sigma$ & Error & MSE & $\sigma$ \\ \hline
200 & 0.304&0.092&0.075 & 0.531&0.282&0.111 & 0.363&0.132&0.115
 &0.685&0.470&0.098 \\

400 & 0.312&0.097&0.099& 0.528&0.279&0.088 & 0.353&0.125&0.112& 0.700&0.490&0.091 \\

600 &0.287&0.082&0.095 & 0.528&0.278&0.089 & 0.699&0.489&0.101 &0.707&0.5&0.089 \\

800 & 0.309&0.096&0.088 & 0.540&0.292&0.090 & 0.361&0.130&0.106
&0.702&0.493&0.098 \\

1000 & \textbf{0.278}&0.077&0.082
 & \textbf{0.516}&0.266&0.086
&\textbf{0.337}&0.113&0.122
 & \textbf{0.684}&0.469&0.090
\\

1200 & 0.298&0.089&0.096 & 0.526&0.277&0.094 & 0.357&0.128&0.112
 & 0.702&0.493&0.091 \\

1400 & 0.31&0.096&0.1 & 0.534&0.285&0.097 & 0.346&0.120&0.098
 &0.697&0.486&0.094 \\

1600 &0.287&0.082&0.094& 0.520&0.271&0.087& 0.344&0.118&0.109
&0.702&0.493&0.099 \\

1800 & 0.282&0.08&0.099 &0.528&0.279&0.093 & 0.354&0.125&0.098
 & 0.708&0.501&0.091\\

2000 &0.292&0.086&0.092
 &0.532&0.283&0.091
 & 0.340&0.115&0.106
 & 0.691&0.478&0.098 \\
\hline
\end{tabular} \\
\vspace*{0.3 cm}

\begin{tabular}{|c|c|c|c|c|c|c|c|c|c|c|c|c|}
\hline
\multirow{ 3}{*}{Particles} & \multicolumn{ 12}{c|}{Distance (m)} \\\cline{2-13} 
&\multicolumn{ 3}{c|}{2.5}  & \multicolumn{ 3}{c|}{3} & \multicolumn{ 3}{c|}{3.5} & \multicolumn{ 3}{c|}{4} \\ \cline{2-13}  
& Error & MSE & $\sigma$ & Error & MSE & $\sigma$& Error & MSE & $\sigma$ & Error & MSE & $\sigma$ \\ \hline
200 & 1.476&2.180&0.019& \textbf{0.909}&0.828&0.022 & 0.973&0.947&0.099 &0.735&0.540&0.049 \\

400 & 1.478&2.185&0.021& 0.943&0.89&0.041 & 0.994&0.989&0.084 &\textbf{0.614}&0.377&0.054
 \\

600 & 1.483&2.201&0.014 & 0.957&0.916&0.032 & \textbf{0.960}&0.923&0.087 & 0.656&0.431&0.067 \\

800 & 1.473&2.170&0.020 & 0.952&0.907&0.041 & 0.992&0.985&0.097& 0.770&0.593&0.033 \\

1000 &1.481&2.194&0.016&0.959&0.921&0.035 &0.972&0.946&0.098 &0.622&0.387&0.081
 \\

1200 & 1.481&2.195&0.017& 0.952&0.907&0.043 & 0.988&0.976&0.114 & 0.803&0.645&0.032 \\

1400 &1.480&2.191&0.019& 0.957&0.916&0.043 & 0.981&0.963&0.089 & 0.640&0.410&0.091 \\

1600 & \textbf{1.478}&2.186&0.015& 0.951&0.906&0.041 & 0.990&0.980&0.095 & 0.689&0.475&0.061 \\

1800 &1.482&2.196&0.018& 0.960&0.923&0.040 & 0.980&0.961&0.096 & 0.797&0.635&0.030 \\

2000 & 1.484&2.204&0.015 & 0.964&0.929&0.037 & 0.967&0.935&0.102 & 0.763&0.582&0.038 \\ 
\hline
\end{tabular}%
\caption{Experimental results (in meters) for different number of particles at the indoor environment.}
\label{table_par}
\end{table*}

The distance estimation results are shown in Fig.~\ref{distance_all}, with the indoor results in Fig.~\ref{indoor_distanceplot} and  the outdoor in Fig.\ref{outdoor_distanceplot}. The solid line depicts the true distance while the averaged and particle filtered measurements are shown as single points and stars with error bars respectively.

    \begin{figure}
\centering
\captionsetup[subfloat]{farskip=4pt}%
\subfloat[Indoor distance estimation.]{\includegraphics[width=0.85\linewidth]{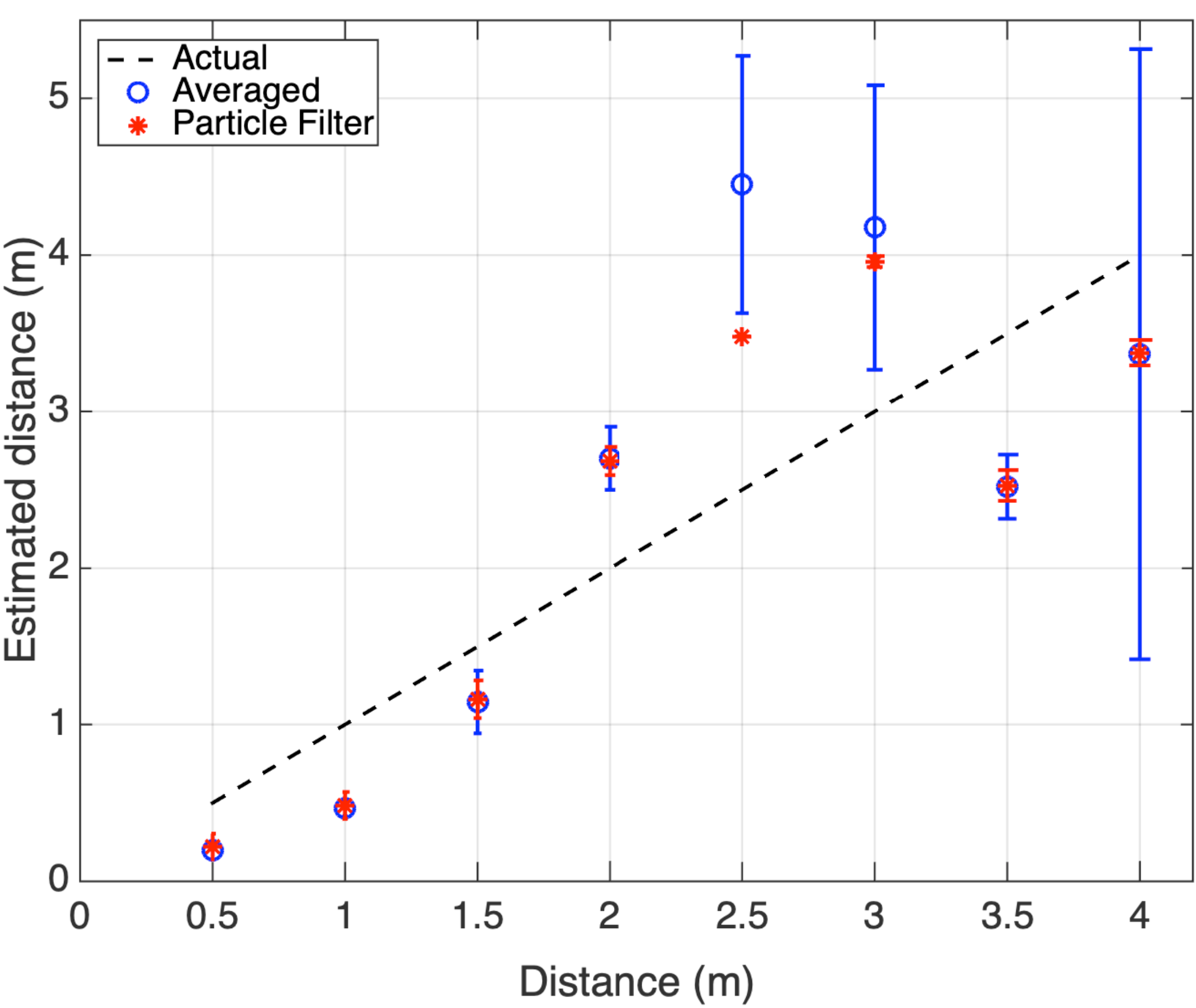}
\label{indoor_distanceplot}}

\subfloat[Outdoor distance estimation.]{\includegraphics[width=0.85\linewidth]{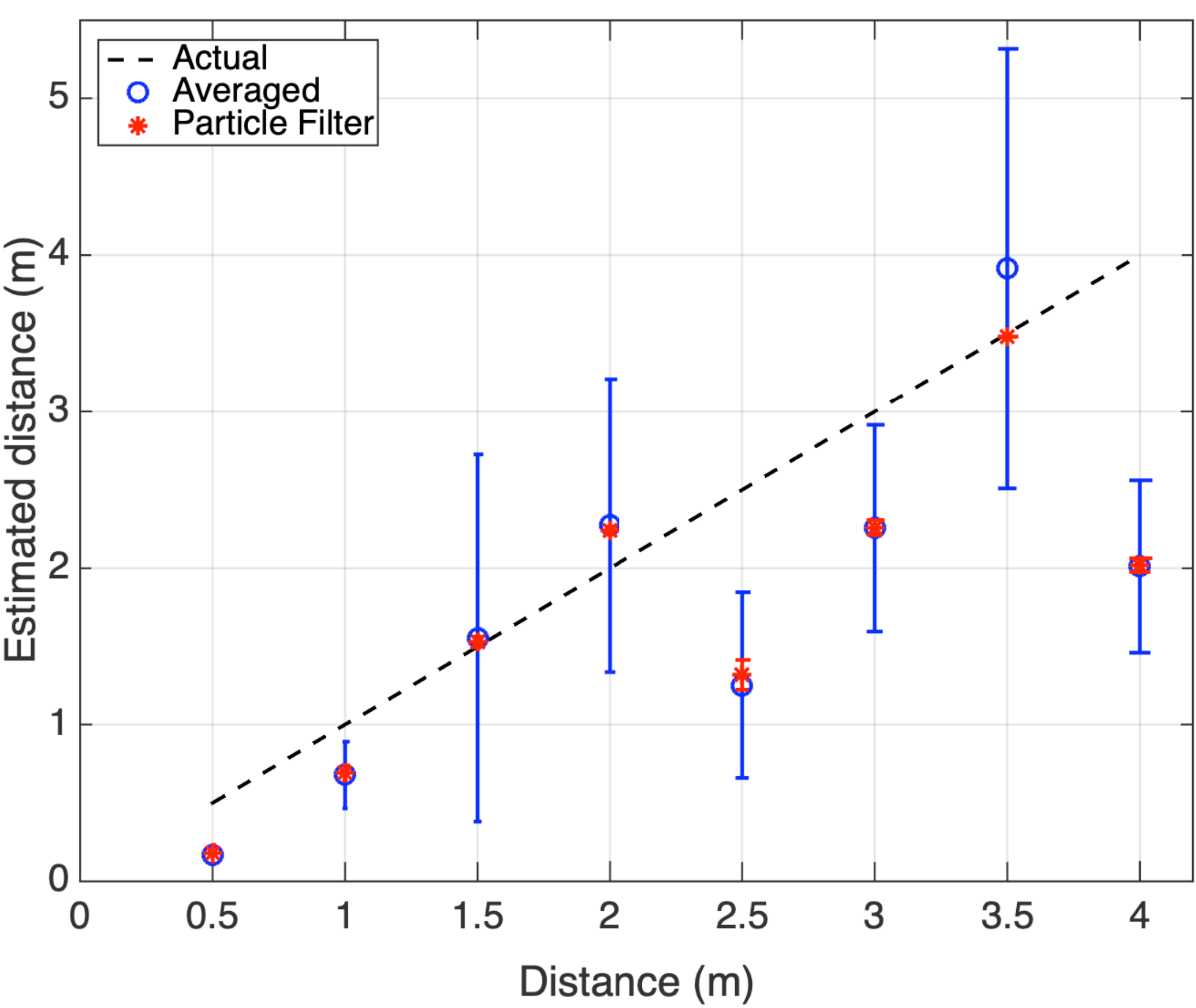}
\label{outdoor_distanceplot}}
\caption{Distance estimation at the two environments.}
\label{distance_all}
\end{figure}

It is clear that the distance estimation is more accurate within 2~m of the beacon than the distance estimations after 2~m. Also, note that the error bars of the particle filtered results are almost unnoticeable. This is due to the fact that the particle filter employs a weighted standard deviation calculation rather than the traditional standard deviation, so, extreme outliers have little influence on the final value and its confidence. 

Overall, the particle filter only shows marginal improvement over the standard averaged data. This is because it is reliant on relatively accurate measurements to be able to converge to the true value. If most of the measurements at a particular distance are wrong, the particle filter fails at estimating the true distance. However, it is better than the use of raw data.

\begin{figure}
\centering
\captionsetup[subfloat]{farskip=4pt}%
\subfloat[CDF for indoor distance estimation.]{\includegraphics[width=0.88\linewidth]{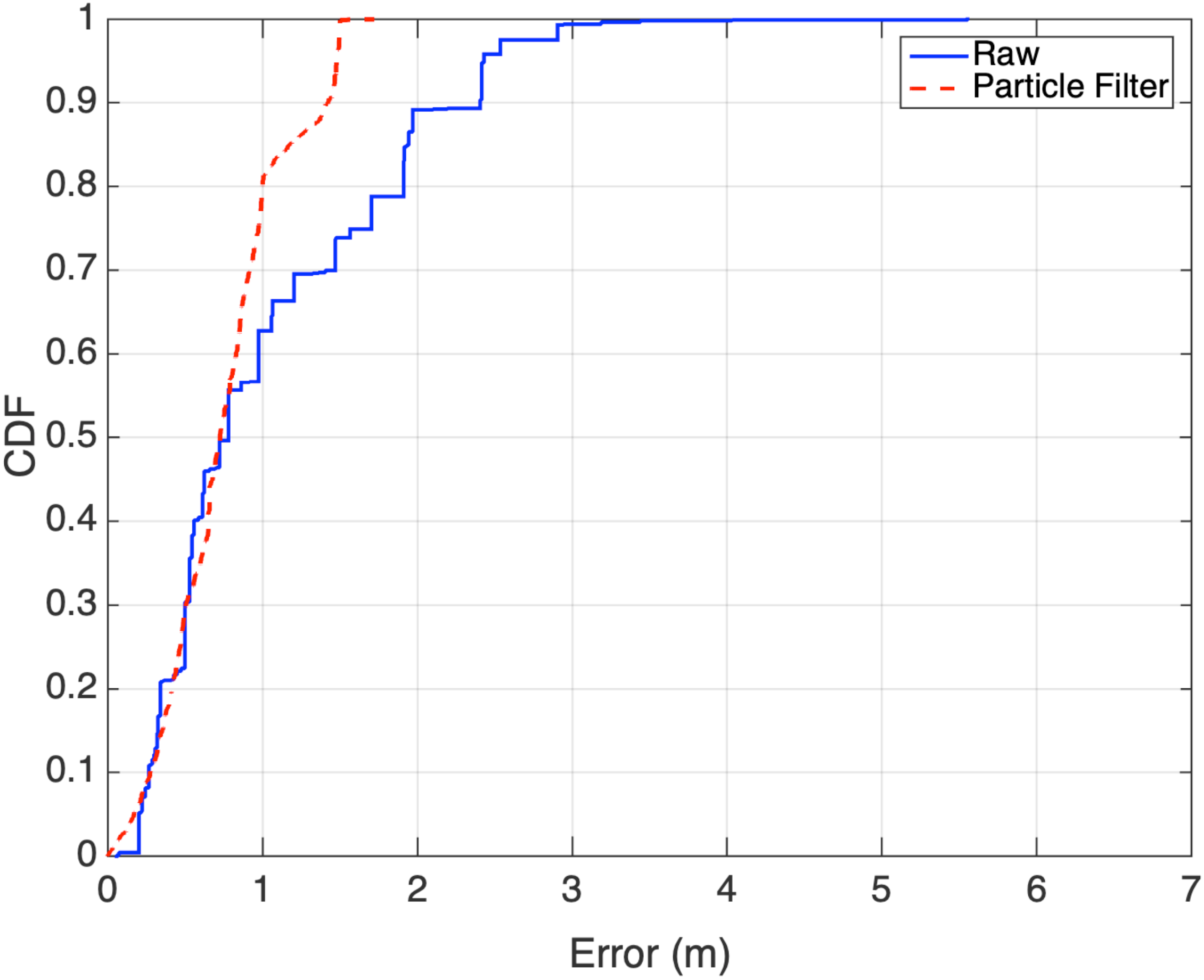}
\label{indoor_cdf}}

\subfloat[CDF for outdoor distance estimation.]{\includegraphics[width=0.88\linewidth]{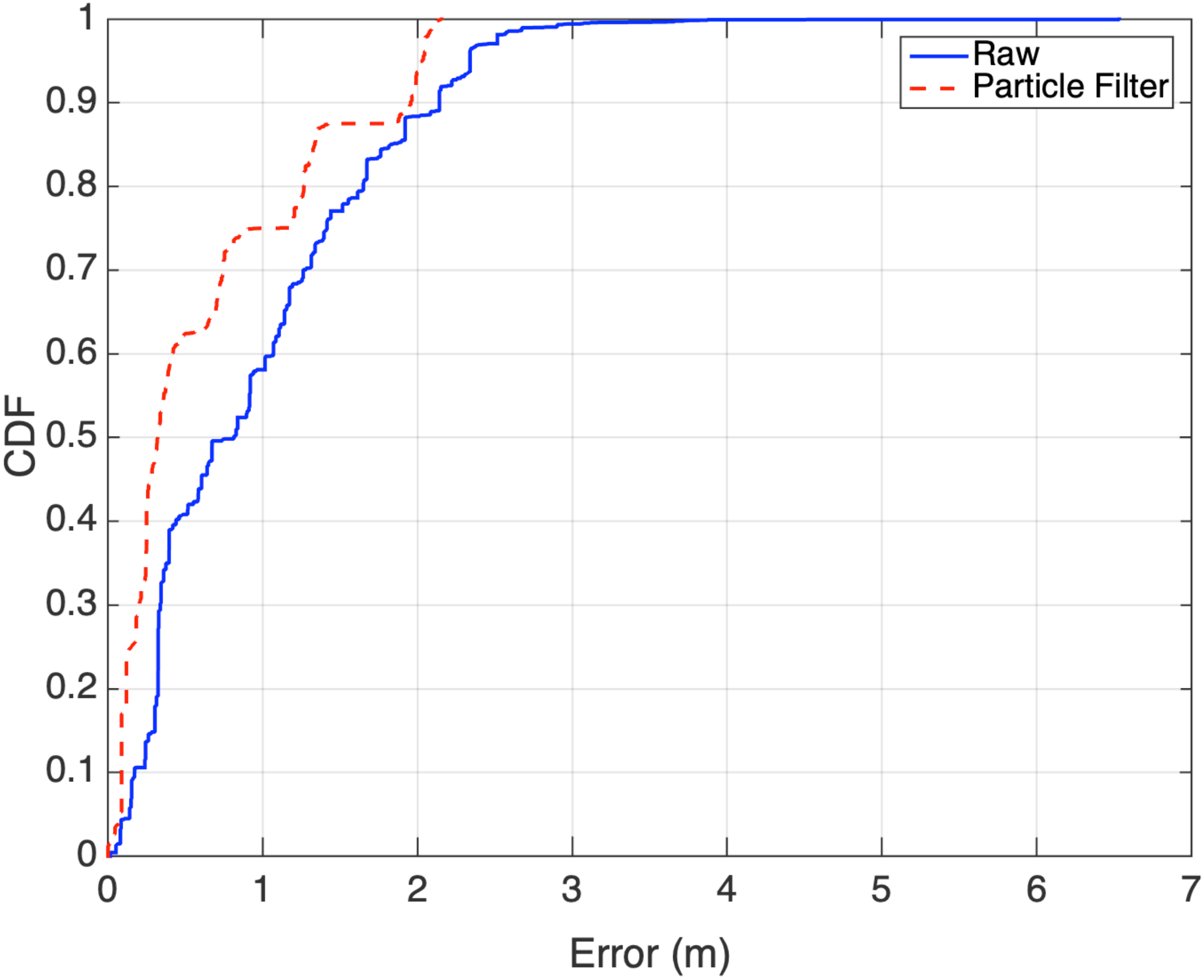}
\label{outdoor_cdf}}
\caption{Cumulative probability error for distance estimation.}
\label{cdf_all}
\end{figure}

The cumulative error for each environment is shown in Fig.~\ref{cdf_all}. For the indoor environment, the error is less than 2.5~m, for 95\% of the time, when raw data are used, while the usage of particle filter reduces the error to 1.5~m, for 95\% of the time, as shown in Fig.\ref{indoor_cdf}. Similarly, for the outdoor environment, the error with the raw data is 2.8~m for 95\% of the time, while it drops to 2~m, for 95\% of the time, when particle filter is used, shown in Fig.~\ref{outdoor_cdf}. According to the experimental results, the use of particle filter can reduce the estimation error by almost 1~m in both environments.

The important interpolation to be gained from this experiment is that there may be reasonable accuracy in locating your vehicle/spot via BLE beacon, with increasing accuracy and confidence as you approach the desired location. This is the base of the guidance capabilities proposed in this system. This experiment also sets the conditions for the following third experiment.

\begin{figure}[t!]
\centering
\captionsetup[subfloat]{farskip=4pt}%
\subfloat[Indoor experiment layout.]{\includegraphics[width=0.7\linewidth]{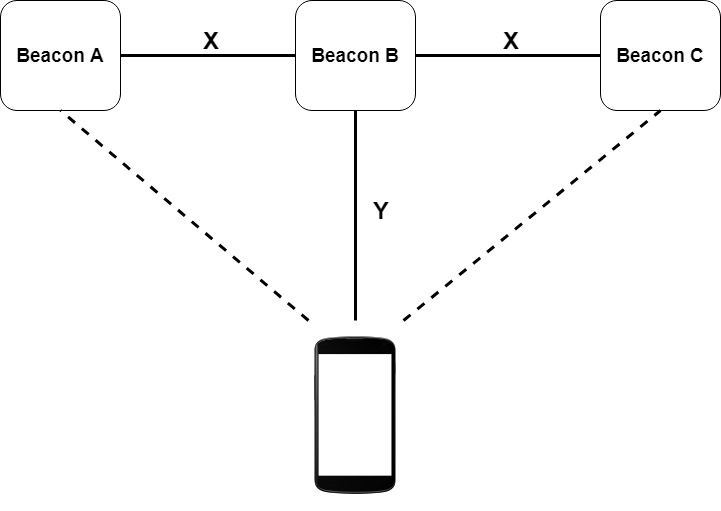}
\label{exp3Layout}}

\subfloat[Parking lot experiment.]{\includegraphics[width=0.8\linewidth]{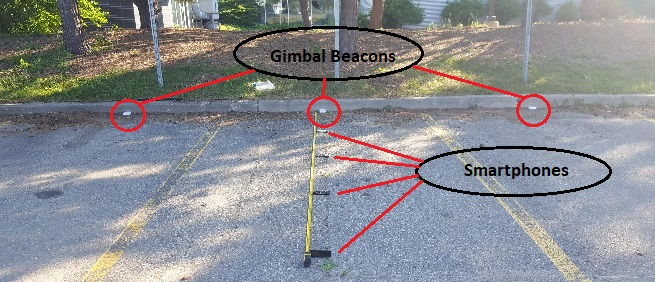}
\label{realpark_out}}
\caption{Proximity identification experiment layout.}
\label{prox_all}
\end{figure}

\subsection{Proximity identification}
This experiment includes the use of three beacons and three associated parking spaces, as shown in Fig.~\ref{prox_all}. Each parking space is given a unique beacon, named A, B, and C. The smartphone collected RSSI values for approximately five minutes, always in straight line with beacon B, but at different distances. The experiment was repeated three times both indoor and outdoor under different conditions, such as the number of cars/ people around as well as weather conditions, for the outdoor experiment. The goal of this experiment is  to determine the maximum distance between the beacons as well as the maximum distance between the beacon and the smartphone, in order for the user to accurately register their vehicle to the desired parking space. This is a critical aspect of the system functionality, since the driver should be able to register the car to the correct parking spot.

Before testing it in an outdoor parking lot, the introduced system is first tested indoors in order to take advantage of the consistent test environment as well as verify experimental feasibility. The layout of the indoor experiment is shown in Fig.~\ref{exp3Layout}. All RSSI estimates from all the three beacons are recorded over a five minute period. From this data, the average RSSI and particle filtered RSSI values are calculated. The greater the RSSI, the closer the smartphone is to the beacon. Based on Pythagorean theorem, the true distance from the smartphone to B will always be closer than to beacons A and C. Hence, the desired result is to have a greater RSSI reading from beacon B, and an equivalent but lower RSSI reading from beacons A and C. In this way, beacon B and its associated spot is chosen, meaning the identification prediction is correct. 
	 
Following the results of the second experiment, we know that the beacon distance estimation is only consistently accurate within the first 2~m. As such, we expect correct predictions within this distance. First, the smartphone is placed at five positions from BLE beacon B. These distances are described by $Y$, where $Y$ =
$\{$ 0.5~m, 1.0~m, 1.5~m, 2.0~m, 2.5~m$\}$. The distance among the three beacons is described by $X$, where $X$~=~$\{$1.0~m, 1.5~m, 2.0~m, 2.5~m, 3.0~m$\}$. This is done in order to verify how robust and consistent the system performs. 
	
	\begin{table}[t]
\centering
\resizebox{\columnwidth}{!}{\begin{tabular}{|c|c|c|c|c|c|c|c|c|c|}
\hline
\multirow{2}{*}{\textbf{X(m)}}       & \multirow{2}{*}{\textbf{Y(m)}} &  \multicolumn{4}{c|}{ \textbf{Raw Data}  }     & \multicolumn{4}{c|}{ \textbf{Particle Filter}  }        \\  \cline{3-10} 
& & A&B&C& \textit{Accuracy (\%)}& A&B&C& \textit{Accuracy (\%)}  \\ \hline
      & 0.5             & 0 &109 &31 & \textbf{77.8} & 0 &111 &29 &79.2 \\ \cline{3-10} 
      & 1               & 1 &72 &20&77.4& 0 &83 &10 &89.2 \\ \cline{3-10}  
      & 1.5             & 44 & 97 &0 &68.7 & 9 &132 &0 &\textbf{93.6}\\ \cline{3-10} 
      & 2               &  26 &18 &40 &21.4& 38 &0 &46 &0\\ \cline{3-10} 
\multirow{-5}{*}{1}   & 2.5             & 20 &1 &115 &0.7 &0 &0 &136&0\\ \hline \hline
     & 0.5             &0 &115&2&98.2&0 &107&10&91.4\\ \cline{3-10} 
      & 1               &1 &53&35&59.5&0 &76&13&85.3
 \\ \cline{3-10}  
      & 1.5             &0 &66&1&\textbf{98.5}&0 &64&3&95.5\\ \cline{3-10} 
      & 2               
&26 &41&8&54.6&0 &75&0&\textbf{100}\\ \cline{3-10} 
\multirow{-5}{*}{1.5}   & 2.5             &65 &3&2&42.8&70 &0&0&0\\ \hline\hline

     & 0.5             &0 &116&0&\textbf{100}&0 &116&0&\textbf{100}\\ \cline{3-10} 
      & 1               &0 &79&5&94&0 &84&0&\textbf{100}
 \\ \cline{3-10}  
      & 1.5            &0 &48&70&40.6&0 &118&0&\textbf{100}\\ \cline{3-10} 
      & 2               
&9 &91&3&88.3&0 &103&0&\textbf{100}\\ \cline{3-10} 
\multirow{-5}{*}{2}   & 2.5&1 &1&112&0.8&0 &0&114&0\\ \hline
\hline

     & 0.5             &0 &121&1&\textbf{99.1}&0 &122&0&\textbf{100}
\\ \cline{3-10} 
      & 1               &41 &30&24&31.5&1 &94&0& 98.9

 \\ \cline{3-10}  
      & 1.5            &2 &123&0&98.4&0 &125&0&\textbf{100}
\\ \cline{3-10} 
      & 2               &1 &120&2&97.5&0 &123&0&\textbf{100}
\\ \cline{3-10} 
\multirow{-5}{*}{2.5}   &2.5&28 &0&93&0&0 &0&121&0
\\ \hline \hline

   & 0.5             &0 &120&0&\textbf{100}&0 &120&0&\textbf{100}
\\ \cline{3-10} 
      & 1               &28 &77&5&70&28 &82&0&74.5

 \\ \cline{3-10}  
      & 1.5            &0 &88&35&71.5&0 &123&0&\textbf{100}

\\ \cline{3-10} 
      & 2               &17 &57&2&75&5 &71&0&93.4
\\ \cline{3-10} 
\multirow{-5}{*}{3}   &2.5&0 &19&101&15.8&0 &0&120&0

\\ \hline

\end{tabular}
}
\caption{Indoor Prediction Results.}
\label{indoor_predTable}
\end{table}

The results for the indoor experiment are shown in  Table~\ref{indoor_predTable}. In general,  as the distance between the beacons increases, the accuracy of the system increases as well. The further away nearby beacons are from each other, the less interference and fewer mistakes they can create to the proposed system. In comparison between the raw data and the particle filter data, the particle filter improves the system performance. There are only two instances, when the distance between the beacons is 1.5~m, that the raw data provide higher accuracy than the filtered data, however, the accuracy of the filtered data is also high. Another interesting insight is that when the accuracy of the raw data is really poor, the particle filter has no accuracy at all. This is expected based on the way particle filters work.
 
 Within the same distance between the beacons, as the smartphone is moving further away from the beacon up to 2~m, the accuracy is great. After the 2~m, the performance degrades, as it was expected from the previous experiment. When the smartphone is too close to the beacon, within 0.5~m or 1.5~m away, which is an acceptable distance for the proposed system, the performance of the system is excellent. Overall, when the distance between the beacons in greater or equal with 2~m and particle filter is used, the system has almost perfect accuracy no matter how close the user is from the beacon.

The results for the outdoor experiment are shown in Table~\ref{outdoor_predTable}. Unlike the indoor experiments, there is only one value for beacon separation, X = 2.7~m, since the parking spaces have a pre-defined width, and the beacons are placed in the middle of each parking spot. Since the distance between the nearby beacons is high, both approaches have high accuracy. In comparison, particle filtered data provide better accuracy for the parking system. For up to two meters distance between the smartphone and the beacon, the accuracy of the system is perfect.

\begin{table}[t!]
\centering
\resizebox{\columnwidth}{!}{\begin{tabular}{|c|c|c|c|c|c|c|c|c|c|}
\hline
\multirow{2}{*}{\textbf{X(m)}}       & \multirow{2}{*}{\textbf{Y(m)}} &  \multicolumn{4}{c|}{ \textbf{Raw Data}  }     & \multicolumn{4}{c|}{ \textbf{Particle Filter}  }        \\  \cline{3-10} 
& & A&B&C& \textit{Accuracy (\%)}& A&B&C& \textit{Accuracy (\%)}  \\ \hline
      & 0.5            &0 &128&0&\textbf{100}&0 &128&0&\textbf{100}

 \\ \cline{3-10} 
      & 1               &0 &126&0&\textbf{100}&0 &126&0&\textbf{100}
 \\ \cline{3-10}  
      & 1.5           &2 &146&17&88.4&0 &165&0&\textbf{100}

\\ \cline{3-10} 
      & 2               &2 &79&12&84.9&0 &93&0&\textbf{100}

\\ \cline{3-10} 
\multirow{-5}{*}{2.7}   & 2.5             &31 &129&1&80.1&18&141&2&87.5
\\ \hline 
\end{tabular}
}
\caption{Outdoor Prediction Results.}
\label{outdoor_predTable}
\end{table}

It is interesting to mention that the system could achieve high and sometimes perfect accuracy within 10 seconds of the user standing in the parking space. It can be inferred that it is not necessary for the user to stay for five minutes to get the correct estimation. The five minutes bound was selected only for experimentation purposes. However, for the particle filter to work, a few samples are necessary.

\subsection{Discussion}

The first experiment is a necessary component of designing the smart parking system. Every environment is different and the measured RSSI of the beacons will change drastically between them, hence, a path loss model should be designed before deploying the system in a given environment. This helps in the appropriate beacon configuration in order to have an optimal distance estimation. As more path loss models are built, we can learn more about system performance and make better tweaks to model parameters, especially the particle filter.  In the distance measurement experiments, it was shown that the most accurate measurements  are within the first 2~m. The beacons increasing their accuracy the closer the user is to them.  

A key aspect of the system is being able to push the correct unique URL, via Eddystone protocol, to the user. The URL brings the user to a secure web server to pay/ register for their unique parking spot. For this to be successful and according to our experiments, the user must remain in their parking spot and should be within 2~m of the beacon that is associated with their spot, as shown in the third experiment. 

Finally, the placement of the beacon is critical. It should be placed at the back center of the spot, but it may also be suitable to place the beacon in the middle of the spot, on the ceiling. All experiments proved that this was successful in this particular test environment. Assuming a good path loss model is developed for alternative environments, similar results should be expected.

\subsubsection{Constraints and potential improvements}
 The extensive experimentation helped to identify some of the system constraints and potential improvements. One major constraint is the internet connectivity. The user needs to communicate with the server to register the parking spot and make the payment. Nowadays, most parking lots can offer internet connection to the user, however, this is something the system administrator needs to include in the design. 

Another issue is the proper placement of the beacons. During experimentation, beacons reported different values when obstacles were between them and the smartphone. As long as all of them experience the same obstacle, such as protecting box, or the obstacle is moving, such as people walking around, there is no problem. However, the system administrator should spend enough time on the proper placement of the beacons to avoid detection errors.

Finally, the presented results are under certain smartphone and beacon placement. Further experimentation with different positioning of the smartphone, such as inside the pocket of the driver or at the passenger seat, is needed to examine the system accuracy.

\section{Conclusions}\label{conclusion}
This paper presents a novel framework for a  smart parking system using BLE beacon devices. The goal was to develop a smartphone application that people can utilize to securely and easily find and pay for parking, while also providing management capabilities for the parking facility owners. 

The introduced system is developed by exploiting BLE beacon technology, building an RSSI path loss model for the desired parking region, and then implementing the RSSI-based distance estimation on the smartphone. Accuracy improvements were made with the implementation of a particle filter. The application is also integrated with cloud-based management services. According to experimental results, the system has sufficient accuracy in terms of parking availability estimation.

\bibliographystyle{IEEEtran}
\bibliography{IEEEabrv,SmartParkv1}

\begin{IEEEbiography}
[{\includegraphics[width=1in,height=1.25in,clip,keepaspectratio]{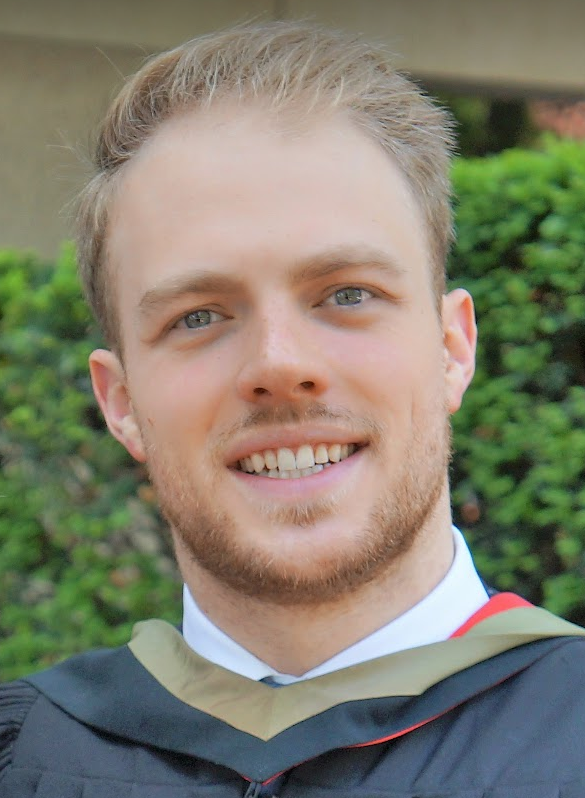}}]{Andrew Mackey} (S'17) received the B.E. degree and the M.A.Sc. degree both in computer engineering from the School of Engineering at the University of Guelph. His research interests lie in the area of Internet of Things with a focus on proximity sensing and indoor localization.
\end{IEEEbiography}

\begin{IEEEbiography}
[{\includegraphics[width=1in,height=1.25in,clip,keepaspectratio]{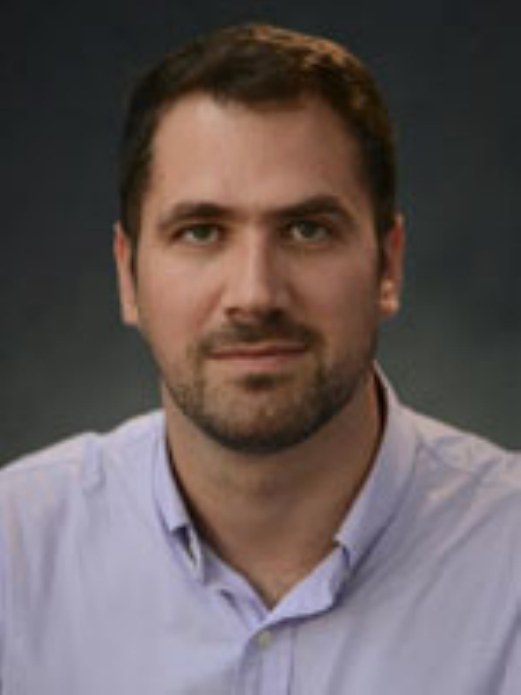}}]{Petros Spachos} (M'14--SM'18) received the Diploma  degree in Electronic and Computer Engineering from the Technical University of Crete, Greece, in 2008, and the M.A.Sc. degree in 2010 and the Ph.D. degree in 2014, both in Electrical and Computer Engineering from the University of Toronto, Canada.  He was a post-doctoral researcher at University of Toronto from September 2014 to July 2015. He is currently an Assistant Professor in the School of Engineering, University of Guelph, Canada. His research interests include experimental wireless networking and mobile computing with a focus on wireless sensor networks, smart cities, and the Internet of Things. He is a Senior Member of the IEEE.
\end{IEEEbiography}

\begin{IEEEbiography}[{\includegraphics[width=1in,height=1.25in,clip,keepaspectratio]{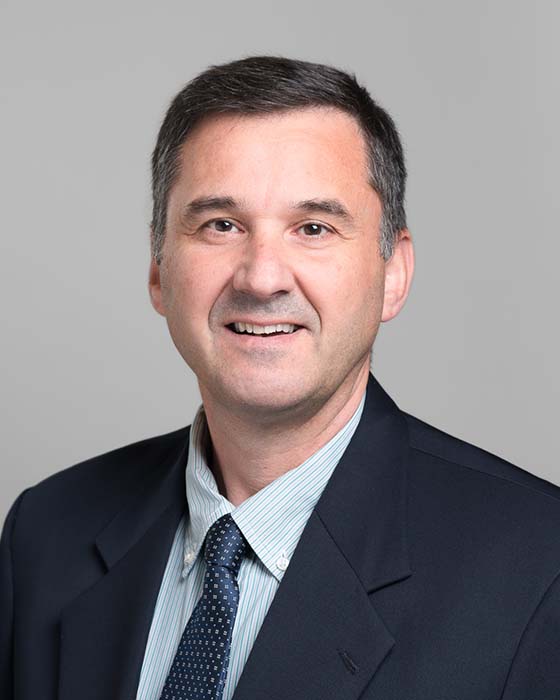}}]{Konstantinos N. (Kostas) Plataniotis}  (S'90--M'92--SM'03--F'12) received his B. Eng. degree in Computer Engineering from University of Patras, Greece and his M.S. and Ph.D. degrees in Electrical Engineering from Florida Institute of Technology Melbourne, Florida. Dr. Plataniotis is currently a Professor with The Edward S. Rogers Sr. Department of Electrical and Computer Engineering at the University of Toronto in Toronto, Ontario, Canada, where he directs the Multimedia Laboratory. He holds the Bell Canada Endowed Chair in Multimedia since 2014. His research interests are primarily in the areas of image/signal processing, machine learning and adaptive learning systems, visual data analysis, multimedia and knowledge media, and affective computing. Dr. Plataniotis is a Fellow of IEEE, Fellow of the Engineering Institute of Canada, and registered professional engineer in Ontario.

Dr. Plataniotis has served as the Editor-in-Chief of the IEEE Signal Processing Letters. He was the Technical Co-Chair of the IEEE 2013 International Conference in Acoustics, Speech and Signal Processing, and he served as the inaugural IEEE Signal Processing Society Vice President for Membership (2014 -2016) and General Co-Chair for the 2017 IEEE GLOBALSIP. He serves as the 2018 IEEE International Conference on Image Processing (ICIP 2018) and the 2021 IEEE International Conference on Acoustics, Speech and Signal Processing (ICASSP 2021) General Co-Chair.
\end{IEEEbiography}

\end{NoHyper}
\end{document}